\documentclass[
    12pt,               % font size
    a4paper,            % paper size
    bibtotoc,           % bibliography in table of contents
    pointlessnumbers,   % heading numbers without point
    ngerman,            % language package
    titlepage,          % separate title page
    leqno,              % formula tags left aligned
    ]{scrarticle}       % document class
\usepackage[T1]{fontenc}
\usepackage{float}
\usepackage[utf8]{inputenc}
\usepackage[ngerman,english]{babel}
\usepackage{url}
\usepackage{graphicx}       % insert pictures
\usepackage[a4paper]{geometry}
\usepackage{amssymb}        % Math symbols
\usepackage{amsmath}        % more Math
\usepackage{amsthm}         % theorem styles
\usepackage{thmtools}       % more theorem tools
\usepackage{csquotes}       % sources
\usepackage{setspace}       % line spacing
\usepackage{tabularx}       % tables
\usepackage{makecell}       % table cells
\usepackage{caption}        % captions
\usepackage{hyperref}
\hypersetup{colorlinks,allcolors=black}
\usepackage[figure]{hypcap}
\usepackage{blindtext}          % Lorem Ipsum plugin
\usepackage{enumitem}           % enumerations
\usepackage{tikz}               % awesome diagrams
\usetikzlibrary{shapes}         % tikz diagram library
\usetikzlibrary{arrows.meta}    % tikz arrows library

\usepackage[style=numeric-comp,maxcitenames=2,maxbibnames=99]{biblatex}
\usepackage[normalem]{ulem}
\useunder{\uline}{\ul}{}

\usepackage{chngcntr}
\counterwithout{figure}{section}
\counterwithout{table}{section}

\RedeclareSectionCommand[%
    beforeskip=0.1\baselineskip,
    afterskip=0.01\baselineskip
]{section}
\RedeclareSectionCommand[%
    beforeskip=0.1\baselineskip,
    afterskip=0.01\baselineskip
]{subsection}
\RedeclareSectionCommand[%
    beforeskip=0.1\baselineskip,
    afterskip=0.01\baselineskip
]{subsubsection}
\RedeclareSectionCommand[%
    beforeskip=0.01\baselineskip,
    afterskip=0.01\baselineskip
]{paragraph}

\addtokomafont{caption}{\small}

\include{hyphenation}

\KOMAoptions{parskip=full*}

\addtokomafont{disposition}{\rmfamily}

\newcommand\secref{Section~\ref}
\newcommand\figref{Figure~\ref}
\newcommand\tableref{Table~\ref}
\newcommand\thmref{Theorem~\ref}
\newcommand\propref{Proposition~\ref}
\newcommand\lemmaref{Lemma~\ref}
\newcommand\corref{Corollary~\ref}
\newcommand\obsref{Observation~\ref}

\newcommand\exref{Example~\ref}

\declaretheoremstyle[
    spaceabove=-4pt,
    spacebelow=0pt,
    headfont=\bfseries\itshape,
    postheadspace=1em,
    qed=\qedsymbol,
    headpunct={}
]{pfstyle}
\declaretheorem[name={Proof},style=pfstyle,unnumbered]{Proof}

\declaretheoremstyle[
    spaceabove=0pt,
    spacebelow=0pt,
    headfont=\bfseries,
    postheadspace=1em,
    headpunct={}
]{mystyle}
\declaretheorem[name={Theorem},style=mystyle]{Theorem}
\declaretheorem[name={Proposition},style=mystyle]{Proposition}
\declaretheorem[name={Lemma},style=mystyle]{Lemma}
\declaretheorem[name={Corollary},style=mystyle]{Corollary}
\declaretheorem[name={Observation},style=mystyle]{Observation}

\declaretheorem[name={Example},style=mystyle]{Example}

\newcommand{\authorName}{Emil Junker}
\newcommand{\type}{Studienprojekt}
\newcommand{\topic}{Manipulative Attacks and Group Identification}

\newcommand{\spiderweb}{
\path (0:0cm) coordinate (O);

    \foreach \X in {1,...,\D}{
    \draw (\X*\A:0) -- (\X*\A:\R);
}

\foreach \Y in {1,...,\U}{
    \foreach \X in {1,...,\D}{
        \path (\X*\A:\Y*\R/\U) coordinate (D\X-\Y);
        \fill (D\X-\Y) circle (1pt);
    }
    \draw [opacity=0.3] (0:\Y*\R/\U) \foreach \X in {1,...,\D}{
        -- (\X*\A:\Y*\R/\U)
    } -- cycle;
}
}

\newcommand{\spiderweblabels}{
\draw (\A:1*\R/\U) node [right] {\tiny unbounded};
\draw (\A:2*\R/\U) node [right] {\tiny param.};
\draw (\A:3*\R/\U) node [right] {\tiny const.};
\draw (\A:4*\R/\U) node [right] {\tiny $=1$};
}

\begin{document}

%%%%%%%%%%%%%%%%%%%%%%%%%%%%%%%%%%%%%%%%%%%%%%%%%%%%%%%%%%%%%%%%
%%%%%%%%%% title page

\subject{\type}

\title{
\normalfont
% \endgraf\rule{\textwidth}{.4pt}
\begingroup
    \centering
    \linespread{1.5}
    \huge\topic
\endgroup
\linespread{1}
% \endgraf\rule{\textwidth}{.4pt}
}

\author{\\ \\ \\ \\ \authorName}

\date{\normalsize{Humboldt-Universität zu Berlin \\
                  Mathematisch-Naturwissenschaftliche Fakultät \\
                  Institut für Informatik \\
                  Betreuer: Dr.\@ Robert Bredereck
                  \paragraph{} \paragraph{}
                  March 15, 2022
                  }}

\titlehead{\begin{center}
    \includegraphics[scale=1]{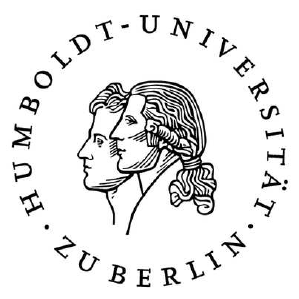}
    \end{center}
}

% line spacing 1.5 from now on
\onehalfspacing

\maketitle

%%%%%%%%%%%%%%%%%%%%%%%%%%%%%%%%%%%%%%%%%%%%%%%%%%%%%%%%%%%%%%%%
%%%%%%%%%% abstract English

\section*{Abstract}

The group identification problem asks to identify a socially qualified subgroup among a group of individuals based on their pairwise valuations. There are several different rules that can be used to determine the social qualification status. In this work, we consider the consent rules, the consensus-start-respecting rule, and the liberal-start-respecting rule.
\\
In the context of group identification, a manipulative attack is the attempt by an outsider to influence the outcome of the selection process through certain means of manipulation. These means include adding, removing, or partitioning individuals, as well as bribing individuals to change their opinion.
\\
In this work, we provide an overview of manipulative attacks in group identification as well as group identification with partial profiles. In particular, we study the computational complexity of the corresponding problems. Most results presented in this work are aggregated from the literature, but we also show results for previously unstudied problems; these include general and exact group control in binary profiles and in ternary profiles, as well as constructive group control in $r$-profiles. For many considered problems, we also study the parameterized complexity.

\thispagestyle{empty}
\clearpage

%%%%%%%%%%%%%%%%%%%%%%%%%%%%%%%%%%%%%%%%%%%%%%%%%%%%%%%%%%%%%%%%
%%%%%%%%%% abstract German

\begin{otherlanguage*}{ngerman}

\section*{Kurzzusammenfassung}

Bei Group Identification geht es darum, aus einer Gruppe von Individuen auf der Grundlage ihrer gegenseitigen Bewertungen eine sozial qualifizierte Untergruppe zu bestimmen. Es gibt verschiedene Regeln, die zur Bestimmung der sozial qualifizierten Individuen verwendet werden können. In dieser Arbeit betrachten wir die \textit{consent rule}, die \textit{consensus-start-respecting rule} und die \textit{liberal-start-respecting rule}.
\\
Im Kontext von Group Identification ist ein manipulativer Angriff der Versuch eines Außenstehenden, das Ergebnis des Auswahlverfahrens durch bestimmte Manipulationsmittel zu beeinflussen. Zu diesen Mitteln gehören das Hinzufügen, Entfernen oder Aufteilen von Individuen sowie die Bestechung von Individuen, ihre Meinung zu ändern.
\\
In dieser Arbeit geben wir einen Überblick über manipulative Angriffe in Group Identification sowie über Group Identification mit \textit{partial profiles}. Insbesondere untersuchen wir die Berechnungskomplexität der entsprechenden Probleme. Die meisten Ergebnisse in dieser Arbeit sind aus der Literatur, aber wir zeigen auch Ergebnisse für bisher nicht untersuchte Probleme; dazu zählen \textit{general} und \textit{exact group control} in \textit{binary profiles} und in \textit{ternary profiles} sowie \textit{constructive group control} in \textit{$r$-profiles}. Für viele der betrachteten Probleme untersuchen wir auch die parametrisierte Komplexität.

\end{otherlanguage*}

\thispagestyle{empty}
\clearpage

%%%%%%%%%%%%%%%%%%%%%%%%%%%%%%%%%%%%%%%%%%%%%%%%%%%%%%%%%%%%%%%%
%%%%%%%%%% table of contents
\newpage

\tableofcontents{}

\setcounter{page}{1}
\pagenumbering{arabic}

\clearpage

%%%%%%%%%%%%%%%%%%%%%%%%%%%%%%%%%%%%%%%%%%%%%%%%%%%%%%%%%%%%%%%%
%%%%%%%%%% content

\section{Introduction}
\label{sec:intro}

There are many situations where one needs to identify a socially qualified subgroup among a group of individuals. For instance, consider a facility whose members want to form a working group for a specific task. In this example, all members have pairwise opinions about whether each member is qualified for the task or not. Another example is when a group of autonomous agents wants to select those among them who are best suited to perform a particular job. Again, the selection in this example is based on the mutual valuations of the agents. To deal with scenarios like these, the group identification model has been developed.

In the group identification model, we are given a set of individuals, each of whom has an opinion about which individuals are qualified in a certain way and which are not. Crucially, each individual also has an opinion about themselves\footnote{Throughout this work, we use the singular ``they'' as a gender-neutral pronoun.}. To determine the subgroup of socially qualified individuals based on the stated opinions, a so-called \textit{social rule} is applied.

The first work on group identification was published by \textcite{K93}, followed by \textcite{KR97}. Since then, a number of different social rules have been proposed and axiomatically characterized \cite{SS03,DSX07,M08,N07}. Also, the problem has been further studied from an economic point of view; see \textcite{D11} for a survey.

In the literature, the consent rules, the consensus-start-respecting (CSR) rule, and the liberal-start-respecting (LSR) rule have received particular attention. A consent rule is characterized by two parameters $s$ and $t$. Under this rule, an individual who qualifies themselves is socially qualified if and only if at least $s$ individuals qualify them; and an individual who disqualifies (i.e.\@ not qualifies) themselves is socially disqualified if and only if at least $t$ individuals disqualify them. With the CSR and LSR rules, the socially qualified individuals are determined iteratively: Initially, the CSR rule considers all individuals who are qualified by everyone as socially qualified; and the LSR rule considers all individuals who qualify themselves as socially qualified. Subsequently, in each iteration, all individuals who are qualified by any of the currently socially qualified individuals are added to the subgroup of socially qualified individuals. The iterative process stops when no new individuals can be added.

Group identification has also been studied in a setting where not all valuations are known in advance. In the literature, this setting is referred to as group identification with \textit{partial profiles} \cite{ERY17,R18}.

\subsection{Manipulative Attacks}
\label{sec:intro_attacks}

In this work, besides the question of determining the socially qualified individuals, we mainly focus on manipulative attacks in group identification. In a manipulative attack, an outsider (called the \textit{attacker}) tries to reach a certain strategic objective. There are four kinds of strategic objectives: With a \textit{constructive} objective, the attacker tries to ensure that certain individuals will be socially qualified. With a \textit{destructive} objective, the attacker tries to ensure that certain individuals will be socially disqualified. With an \textit{exact} objective, the attacker tries to precisely specify for each individual whether they should be socially qualified or not. Finally, in the most general setting (also called \textit{constructive+destructive}), the attacker tries to make certain individuals socially qualified and certain other individuals socially disqualified.

There are different means which the attacker may use to reach their objective. In \textit{group control} scenarios, the attacker can add, delete, or partition individuals. In \textit{group bribery} scenarios, the attacker can bribe individuals to change their opinion. Each strategic objective can be combined with each means of manipulation to form a group control (resp.\@ group bribery) problem. For example, consider the \textit{destructive group control by deleting individuals} problem: In this scenario, the attacker is allowed to delete a certain number of individuals from the set, with the goal of making certain individuals in the remaining set socially disqualified.

The precise steps the attacker must take to reach their objective depend on which social rule is used for evaluation. For each combination of a strategic objective, a means of manipulation, and a social rule, we can study the feasibility and the computational complexity of the corresponding problem. Sometimes, it is not possible for the attacker to reach their objective by the given means. In this case, we say that the used social rule is \textit{immune} to the manipulative attack. Otherwise, we say that the social rule is \textit{susceptible} to the manipulative attack, and we then try to classify the problem as either polynomial-time solvable or NP-hard. When the problem is NP-hard, it is possible for the attacker to reach their objective, but computing the solution is generally difficult. This implies that the used social rule offers at least some protection against the manipulative attack.

Throughout this work, we denote the given set of individuals as $N$. We use $A^+$ to denote the constructive target set, i.e.\@ the set of individuals the attacker wants to make socially qualified. Likewise, we use $A^-$ to denote the destructive target set, i.e.\@ the set of individuals the attacker wants to make socially disqualified. Naturally, we require that $A^+ \cap A^- = \emptyset$. For the exact objective, we usually add the premise $A^+ \cup A^- = N$. We now give a brief introduction to the different problem settings. Formal definitions of the problems will be provided in \secref{sec:problem_definitions}.

In the scenario where the attacker may add individuals, there is a predefined ``pool'' of individuals to choose from. Initially, only a subset of individuals $T \subseteq N$ is considered in the group identification process. The attacker can then select a certain number of individuals from $N \setminus T$ to be added to $T$. This can be thought of as motivating additional individuals to participate in the process who otherwise would not have participated.

In the scenario where the attacker may delete individuals, we start with an initial set $N$ and allow the attacker to delete a certain number of individuals from $N$. This can be thought of as preventing some individuals from participating in the selection process. In our model, the attacker is not allowed to delete individuals that are in $A^+$ or $A^-$; this could be justified by the fact that the attacker really wants to make the individuals in $A^-$ socially disqualified (and not simply delete them from the set).

For group control by partitioning, the attacker can organize the evaluation procedure as a two-stage process. In the first stage of the selection, the individuals are split into two partitions $U$ and $N \setminus U$, and the socially qualified individuals are determined separately for both partitions. Subsequently, in the second stage of the selection, only individuals who survived the first stage are considered. Multi-stage selection procedures are common in many real-world scenarios (e.g.\@ think of organizational units or electoral districts).

In the bribery problems, the attacker can bribe the individuals to change their opinion. Each individual has a price of being bribed, and the attacker has a certain budget they are allowed to spend. In this model, we assume that the attacker can arbitrarily alter the outgoing valuations of each bribed individual.

There is also an alternative bribery model called \textit{microbribery} where the attacker needs to pay for each changed valuation separately. In this model, each valuation has a price of being changed, and the attacker again has a certain budget they are allowed to spend.

Finally, for both bribery and microbribery, there also exist unpriced variants where the attacker is allowed to make a certain number of bribes (i.e.\@ the bribes all have unit costs).

The study of manipulative attacks in group identification was initiated by \textcite{YD18}. They considered group control by adding, deleting, and partitioning of individuals with a constructive objective. \textcite{ERY20} extended their work to group control with a destructive objective and to unpriced group bribery. Furthermore, \textcite{EY20} studied the constructive and exact unpriced microbribery problems. Finally, \textcite{BBKL20} were the first to consider group control and bribery problems (including priced bribery and microbribery) with a general and exact objective.

\subsection{Related Work}
\label{sec:intro_related}

In this section, we briefly explore the existing literature on problem settings to which the group identification problem and group control/bribery attacks are formally related.

Group identification is related to multi-winner voting systems \cite{FSST17,EFSS17}. In a voting system, there is a set of voters and a set of candidates, and the voters express preference over the candidates by casting votes. A subset of winning candidates is then selected based on the voting. However, group identification differs from multi-winner voting systems in several aspects: First, in group identification, there is only one set of individuals, i.e.\@ the set of voters and the set of candidates coincide. Second, the subgroup of socially qualified individuals should not necessarily be thought of as ``winners''. Rather, the group identification model can be used to determine all kinds of subgroups, e.g.\@ the conservative-minded members of a political party. Finally, in group identification, the size of the socially qualified subgroup is not fixed. In contrast, voting systems usually have the number of winners fixed in advance, which is why they require a tie-breaking mechanism.

Among all voting systems, group identification is most closely related to approval voting which has been extensively studied \cite{BEHHR10,FB81,HHR07,LS10,L11,YG17}. In approval voting, each voter either approves or disapproves each candidate, and the candidates with the most approvals are declared winners. However, as with most voting systems, the number of winners in approval voting is usually fixed in advance, so a tie-breaking mechanism is required.

There also exist multi-winner voting systems where the number of winners is not fixed \cite{BBMN04,K16,FST20,YW18,LM21,DPZ16}. However, these systems use rules that are different to the social rules studied in this work.

In the setting where the number of winners is fixed, manipulative attacks for (approval-based) multi-winner voting systems have been examined in the literature \cite{AGGMMW15,BKN21,BFNT21,OZE13,KF19,PRZ07,Y19,Y20}. Originally, the study of control scenarios where voters are added, deleted, or partitioned was initiated in the context of elections \cite{BTT92,MPRZ08,YG14,YG15}. Bribery in elections where the preferences of a certain number of voters are changed \cite{FHH09} and microbribery in elections where the attacker pays for each flipped preference \cite{FHHR09} have also been studied. For a survey on control and bribery problems in elections, see \textcite{FR16}.

Bribery in elections is related to the concept of \textit{margin of victory} \cite{C11,MRSW11} and there are several papers studying the connection between these concepts \cite{BBFN21,FST17,X12}. Intuitively, the number of voters whose valuations would need to be changed to turn a certain candidate into a winner (or to turn a winning candidate into a losing one) can be used as a measure for the margin of victory of said candidate. This approach can also be used to determine the robustness of the outcome of an election in general.

The topic of microbribery has also been studied in the context of lobbying where a lobbyist tries to influence voters who are deciding a set of yes/no issues \cite{BEFGMR14,CFRS07}. However, the social rules used in group identification are different from the aggregation procedures used in lobbying problems.

Finally, group identification is also related to peer selection \cite{AFPT11,ALMRW16}. In the peer selection problem, we want to select a subset from a given set of individuals based on their mutual valuations. However, in the peer selection model, each individual only evaluates other individuals, and it is assumed that each individual is willing to be selected. Also, unlike group identification, the size of the subset in peer selection is fixed in advance.

\subsection{Organisation}
\label{sec:intro_organization}

The remainder of this work is organized as follows: In \secref{sec:basic_notation}, we introduce our basic notation. In \secref{sec:social_rules}, we define the relevant social rules, and in \secref{sec:problem_definitions}, we provide formal definitions of the various group control and bribery problems that will be examined in this work. Subsequently, in \secref{sec:binary}, we give an overview of the computational complexity results for manipulative attacks in binary profiles. \secref{sec:r_profiles} is devoted to the restriction of certain problems to $r$-profiles. In \secref{sec:ternary}, we study manipulative attacks in ternary profiles. \secref{sec:partial} is about group identification in partial profiles. In \secref{sec:potential_params}, we briefly discuss potential parameters for the examined problems. \secref{sec:conclusion} concludes this work with a summary of our results and some ideas for future research.

\clearpage

\section{Basic Notation}
\label{sec:basic_notation}

Let $N$ be a set of individuals. In group identification, every individual $a \in N$ has an opinion about who from $N$ is qualified in a certain way and who is not.

A binary profile over $N$ is a function $\varphi : N \times N \rightarrow \{ -1, 1 \}$. For $a,b \in N$, we write $\varphi(a,b) = 1$ when $a$ thinks $b$ is qualified and say \textit{$a$ qualifies $b$}. We write $\varphi(a,b) = -1$ when $a$ thinks $b$ is not qualified and say \textit{$a$ disqualifies $b$}.

Let $N$ be a set of individuals and let $a \in N$. For a profile $\varphi$ over $N$, we sometimes refer to the entries of the form $\varphi(a, \cdot)$ as \textit{outgoing qualifications of $a$} and entries of the form $\varphi(\cdot, a)$ as \textit{incoming qualifications of $a$}. Also, for any $T \subseteq N$ and $x \in \{ -1, 1 \}$, we sometimes write $T^x_\varphi(a)$ to denote the set $\{ a^\prime \in T : \varphi(a^\prime, a) = x \}$. In particular, $N^1_\varphi(a)$ denotes the set of all individuals who qualify $a$, and $N^{-1}_\varphi(a)$ denotes the set of all individuals who disqualify $a$.

For a given set $N$ of individuals and a binary profile $\varphi$ over $N$, we can construct a so-called (directed) \textit{qualification graph} $G_{N, \varphi} = (N, E)$ with $(a, b) \in E$ if and only if $\varphi(a, b) = 1$. Interpreting the problem instances as graphs can sometimes be useful for reductions. Also, we can use $G_{N, \varphi}$ to derive graph parameters for a given instance.

When considering bribery problems, we use price functions to denote the cost of bribing a certain individual to change their opinion. A bribery price function $\rho : N \rightarrow \mathbb{N}$ assigns every individual $a \in N$ a positive integer price. For a set of individuals $N^\prime \subseteq N$, we write $\rho(N^\prime)$ to denote $\sum_{a \in N^\prime} \rho(a)$. A microbribery price function $\rho : N \times N \rightarrow \mathbb{N}$ assigns every pair $(a,b) \in N \times N$ of individuals a positive integer price, corresponding to the cost of bribing $a$ to change their opinion about $b$. For a set of pairs of individuals $M \subseteq N \times N$, we write $\rho(M)$ to denote $\sum_{m \in M} \rho(m)$. When considering unpriced bribery problems [resp.\@ unpriced microbribery problems], we let $\rho$ assign unit prices to all individuals [resp.\@ all pairs of individuals].

\clearpage

\section{Social Rules}
\label{sec:social_rules}

Given a set $N$ of individuals and a profile $\varphi$ over $N$, a social rule is used to determine which individuals are socially qualified and which are not. In this section, we give formal definitions of the social rules that will be studied in this work. We also provide a small example where we list the socially qualified individuals under different social rules.

\subsection{Definitions}

In general, a social rule is a function $f$ that assigns a subset $f(T, \varphi) \subseteq T$ to each pair $(T, \varphi)$ where $T \subseteq N$. We refer to the individuals in $f(T, \varphi)$ [resp.\@ $T \setminus f(T, \varphi)$] as the \textit{socially qualified [resp.\@ socially disqualified] individuals of $T$ with respect to $f$ and $\varphi$}.

Let $N$ be a set of $n$ individuals and $\varphi$ a binary profile over $N$. Below, we provide the definitions for the social rules that will be studied in this work.

\textbf{Consent rules} $f^{(s, t)}$:
A consent rule is specified by two parameters $s, t \in \mathbb{N}$, called the consent quotas. For every subset $T \subseteq N$ and every individual $a \in T$, it holds:
\setlist[itemize,1]{label=\normalfont\bfseries\textendash}
\begin{itemize}[nolistsep,topsep=-4pt]
\item
If $\varphi(a, a) = 1$
then $a \in f^{(s, t)}(T, \varphi)$ if and only if
$|{a^\prime \in T : \varphi(a^\prime, a) = 1}| \geq s$.
\item
If $\varphi(a, a) = -1$
then $a \not\in f^{(s, t)}(T, \varphi)$ if and only if
$|{a^\prime \in T : \varphi(a^\prime, a) = -1}| \geq t$.
\end{itemize}
\setlist[itemize,1]{label=\textbullet}

In other words, an individual who qualifies themselves is socially qualified if and only if at least $s-1$ other individuals qualify them. And an individual who disqualifies themselves is socially disqualified if and only if at least $t-1$ other individuals disqualify them.

The consent rule for $s = t = 1$ is also referred to as \textit{liberal rule} in the literature. Under the liberal rule, an individual is socially qualified if and only if they qualify themselves.

To prevent situations where an individual could manipulate their own social qualification status by changing their opinion about themselves, we require that $s + t \leq n + 2$. This constraint ensures the monotonicity property of the consent rule, i.e.\@ a socially disqualified individual is still socially disqualified when someone who qualifies them turns to disqualify them \cite{SS03}. In particular, if an individual qualifies themselves but is socially disqualified (less than $s$ individuals qualify them), the constraint ensures that they cannot become socially qualified by simply disqualifying themselves (and hoping that less than $t$ individuals disqualify them).

\textbf{Consensus-start-respecting rule} $f^{\text{CSR}}$:
This rule determines the socially qualified individuals iteratively. In a first step, all individuals who are qualified by everyone are added to the set of socially qualified individuals. Then, in each iteration, any individual who is qualified by any of the currently socially qualified individuals is added to the set of socially qualified individuals. Once no new individuals can be added, the iterative process terminates.

More formally, for every $T \subseteq N$, let
$$
K^{\text{C}}_0(T, \varphi) = \{a \in T : \varphi(a^\prime, a) = 1 \text{ for all } a^\prime \in T\}.
$$
Then, for each positive integer $i$, let
$$
K^{\text{C}}_i(T, \varphi) = K^{\text{C}}_{i-1}(T, \varphi) \cup \{a \in T : \exists~ a^\prime \in K^{\text{C}}_{i-1}(T, \varphi) \text{ s.t.\@ } \varphi(a^\prime, a) = 1\}.
$$
We define $f^{\text{CSR}}(T, \varphi) = K^{\text{C}}_i(T, \varphi)$ for the smallest $i$ with $K^{\text{C}}_i(T, \varphi) = K^{\text{C}}_{i-1}(T, \varphi)$.

\textbf{Liberal-start-respecting rule} $f^{\text{LSR}}$:
This rule also determines the socially qualified individuals iteratively. Initially, all individuals who qualify themselves are considered to be socially qualified. Then, in each iteration, any individual who is qualified by any of the currently socially qualified individuals is added to the set of socially qualified individuals. Once no new individuals can be added, the iterative process terminates.

More formally, for every $T \subseteq N$, let
$$
K^{\text{L}}_0(T, \varphi) = \{a \in T : \varphi(a, a) = 1\}.
$$
Then, for each positive integer $i$, let
$$
K^{\text{L}}_i(T, \varphi) = K^{\text{L}}_{i-1}(T, \varphi) \cup \{a \in T : \exists~ a^\prime \in K^{\text{L}}_{i-1}(T, \varphi) \text{ s.t.\@ } \varphi(a^\prime, a) = 1\}.
$$
We define $f^{\text{LSR}}(T, \varphi) = K^{\text{L}}_i(T, \varphi)$ for the smallest $i$ with $K^{\text{L}}_i(T, \varphi) = K^{\text{L}}_{i-1}(T, \varphi)$.

\subsection{Example}

We now present a simple example, using the same setup as \textcite[6]{ERY20}.

\begin{Example} \label{ex:one}
Let $N = \{ a_1, a_2, a_3, a_4, a_5 \}$ be a set of five individuals and $\varphi$ a binary profile over $N$ defined as follows ($\varphi(a_i, a_j)$ is given by the matrix entry in the $i$-th row and $j$-th column):

\capstartfalse % figure has no caption
\begin{figure}[!htb]
\centering
\begin{minipage}[b]{.45\textwidth}

\renewcommand{\arraystretch}{1.2}
\begin{equation*}
\begin{array}{rrrrrr}
    & a_1 & a_2 & a_3 & a_4 & a_5  \\
a_1 &  1 &  1 &  1 & -1 &  1 \\
a_2 & -1 & -1 &  1 & -1 &  1 \\
a_3 & -1 &  1 &  1 & -1 & -1 \\
a_4 &  1 &  1 &  1 &  1 & -1 \\
a_5 & -1 &  1 &  1 & -1 & -1
\end{array}
\end{equation*}
\caption*{Definition matrix of $\varphi$.}

\end{minipage}\hfill
\begin{minipage}[b]{.55\textwidth}
\centering

\raisebox{0.1\height}{
\begin{tikzpicture}[
    > = Stealth,
    shorten > = 1pt,
    auto,
    node distance = 1.8cm,
    main node/.style = {circle,draw}
]

    \node[main node] (1) {$a_1$};
    \node[main node] (2) [above right of=1] {$a_2$};
    \node[main node] (3) [right of=2] {$a_3$};
    \node[main node] (4) [below right of=1] {$a_4$};
    \node[main node] (5) [right of=4] {$a_5$};

    \path[->]
    (1) edge[loop left]         node {} (1)
        edge                    node {} (2)
        edge[bend right=10]     node {} (3)
        edge                    node {} (5)
    (2) edge[bend right=10]     node {} (3)
        edge[bend right=10]     node {} (5)
    (3) edge[bend right=10]     node {} (2)
        edge[loop right]        node {} (3)
    (4) edge                    node {} (1)
        edge                    node {} (2)
        edge                    node {} (3)
        edge[loop left]         node {} (4)
    (5) edge[bend right=10]     node {} (2)
        edge                    node {} (3)
    ;

\end{tikzpicture}
}
\caption*{Corresponding qualification graph $G_{N, \varphi}$.}

\end{minipage}
\end{figure}
\capstarttrue % re-enable figure captions

Using the liberal rule $f^{(1,1)}$, the set of socially qualified individuals is simply $f^{(1,1)}(N, \varphi) = \{ a_1, a_3, a_4 \}$, i.e.\@ everyone who qualifies themselves. Using the consent rule with $s=1$ and $t=2$, the set of socially qualified individuals grows to $f^{(1,2)}(N, \varphi) = \{ a_1, a_2, a_3, a_4 \}$. For the consent rule with $s=2$ and $t=1$, we instead get $f^{(2,1)}(N, \varphi) = \{ a_1, a_3 \}$.

With the consensus-start-respecting rule, we initially have $K^{\text{C}}_0(N, \varphi) = \{ a_3 \}$ as $a_3$ is the only individual who is qualified by everyone. Because $a_3$ qualifies $a_2$, we get $K^{\text{C}}_1(N, \varphi) = \{ a_2, a_3 \}$. Since $a_2$ also qualifies $a_5$, we then get $K^{\text{C}}_2(N, \varphi) = \{ a_2, a_3, a_5 \}$. The individual $a_5$ qualifies no further individuals, so we obtain the final result $f^{\text{CSR}}(N, \varphi) = \{ a_2, a_3, a_5 \}$.

With the liberal-start-respecting rule, we have $K^{\text{L}}_0(N, \varphi) = \{ a_1, a_3, a_4 \}$. Because $a_1$ also qualifies $a_2$ and $a_5$, we then get $K^{\text{L}}_1(N, \varphi) = \{ a_1, a_2, a_3, a_4, a_5 \} = f^{\text{LSR}}(N, \varphi)$.

\end{Example}

\clearpage

\section{Problem Definitions}
\label{sec:problem_definitions}

In this section, we provide formal definitions of the problems outlined in \secref{sec:intro_attacks}.

\bigskip

\begin{tabularx}{\textwidth}{lX}
\hline
\multicolumn{2}{l}{
$f$-\textsc{Group Control by Adding Individuals} (GCAI)
} \\
\hline
\textbf{Given:} &
A 6-tuple $(N, \varphi, A^+, A^-, T, \ell)$ of a set $N$ of individuals, a profile $\varphi$ over $N$, three subsets $A^+, A^-, T \subseteq N$ with $A^+ \cap A^- = \emptyset$ and $A^+, A^- \subseteq T$, and a positive integer $\ell$. \\
\textbf{Question:} &
Is there a subset $U \subseteq N \setminus T$ such that $|U| \leq \ell$ and $A^+ \subseteq f(T \cup U, \varphi)$ and $A^- \cap f(T \cup U, \varphi) = \emptyset$? \\
\hline
\end{tabularx}

In addition to the general $f$-GCAI problem, there are the three special variants
\begin{itemize}
\item
$f$-\textsc{Constructive Group Control by Adding Individuals} (CGCAI) where the set $A^-$ is dropped,
\item
$f$-\textsc{Destructive Group Control by Adding Individuals} (DGCAI) where the set $A^+$ is dropped,
\item
$f$-\textsc{Exact Group Control by Adding Individuals} (EGCAI) where we add the premise $A^+ \cup A^- = T$.
\end{itemize}

\bigskip

\begin{tabularx}{\textwidth}{lX}
\hline
\multicolumn{2}{l}{
$f$-\textsc{Group Control by Deleting Individuals} (GCDI)
} \\
\hline
\textbf{Given:} &
A 5-tuple $(N, \varphi, A^+, A^-, \ell)$ of a set $N$ of individuals, a profile $\varphi$ over $N$, two subsets $A^+, A^- \subseteq N$ with $A^+ \cap A^- = \emptyset$, and a positive integer $\ell$. \\
\textbf{Question:} &
Is there a subset $U \subseteq N \setminus (A^+ \cup A^-)$ such that $|U| \leq \ell$ and $A^+ \subseteq f(N \setminus U, \varphi)$ and $A^- \cap f(N \setminus U, \varphi) = \emptyset$? \\
\hline
\end{tabularx}

In addition to the general $f$-GCDI problem, there are the two special variants
\begin{itemize}
\item
$f$-\textsc{Constructive Group Control by Deleting Individuals} (CGCDI) where the set $A^-$ is dropped,
\item
$f$-\textsc{Destructive Group Control by Deleting Individuals} (DGCDI) where the set $A^+$ is dropped.
\end{itemize}

Note that there is no $f$-\textsc{Exact Group Control by Deleting Individuals} problem. The reason for this is that we only allow the attacker to delete individuals from $N \setminus (A^+ \cup A^-)$. Hence, the attacker would not be allowed to delete anyone when $A^+ \cup A^- = N$.

It could be interesting to consider problem variants where we allow the attacker to also delete individuals from $A^-$. However, we do not do that in this work.

\bigskip

\begin{tabularx}{\textwidth}{lX}
\hline
\multicolumn{2}{l}{
$f$-\textsc{Group Control by Partitioning of Individuals} (GCPI)
} \\
\hline
\textbf{Given:} &
A 4-tuple $(N, \varphi, A^+, A^-)$ of a set $N$ of individuals, a profile $\varphi$ over $N$, and two subsets $A^+, A^- \subseteq N$ with $A^+ \cap A^- = \emptyset$. \\
\textbf{Question:} &
Is there a subset $U \subseteq N$ such that $A^+ \subseteq f(V, \varphi)$ and $A^- \cap f(V, \varphi) = \emptyset$ where $V = f(U, \varphi) \cup f(N \setminus U, \varphi)$? \\
\hline
\end{tabularx}

In addition to the general $f$-GCPI problem, there are the three special variants
\begin{itemize}
\item
$f$-\textsc{Constructive Group Control by Partitioning of Individuals} (CGCPI) where the set $A^-$ is dropped,
\item
$f$-\textsc{Destructive Group Control by Partitioning of Individuals} (DGCPI) where the set $A^+$ is dropped,
\item
$f$-\textsc{Exact Group Control by Partitioning of Individuals} (EGCPI) where we add the premise $A^+ \cup A^- = N$.
\end{itemize}

\bigskip

\begin{tabularx}{\textwidth}{lX}
\hline
\multicolumn{2}{l}{
$f$-\textsc{\$Group Bribery} (\$GB)
} \\
\hline
\textbf{Given:} &
A 6-tuple $(N, \varphi, A^+, A^-, \rho, \ell)$ of a set $N$ of individuals, a profile $\varphi$ over $N$, two subsets $A^+, A^- \subseteq N$ with $A^+ \cap A^- = \emptyset$, a bribery price function $\rho : N \rightarrow \mathbb{N}$, and a positive integer $\ell$. \\
\textbf{Question:} &
Is there a way to obtain a profile $\varphi^\prime$ over $N$ by changing the outgoing qualifications in $\varphi$ of some individuals $U \subseteq N$ such that $\rho(U) \leq \ell$ and $A^+ \subseteq f(N, \varphi^\prime)$ and $A^- \cap f(N, \varphi^\prime) = \emptyset$? \\
\hline
\end{tabularx}

In addition to the general $f$-\$GB problem, there are the three special variants
\begin{itemize}
\item
$f$-\textsc{\$Constructive Group Bribery} (\$CGB) where the set $A^-$ is dropped,
\item
$f$-\textsc{\$Destructive Group Bribery} (\$DGB) where the set $A^+$ is dropped,
\item
$f$-\textsc{\$Exact Group Bribery} (\$EGB) where we add the premise $A^+ \cup A^- = N$.
\end{itemize}

Also, there are the four unpriced versions $f$-\textsc{Group Bribery} (GB), $f$-\textsc{Constructive Group Bribery} (CGB), $f$-\textsc{Destructive Group Bribery} (DGB), and $f$-\textsc{Exact Group Bribery} (EGB) where $\rho$ assigns price $1$ to every individual, i.e.\@ $\rho(a) = 1$ for all $a \in N$.

\bigskip

\begin{tabularx}{\textwidth}{lX}
\hline
\multicolumn{2}{l}{
$f$-\textsc{\$Group Microbribery} (\$GMB)
} \\
\hline
\textbf{Given:} &
A 6-tuple $(N, \varphi, A^+, A^-, \rho, \ell)$ of a set $N$ of individuals, a profile $\varphi$ over $N$, two subsets $A^+, A^- \subseteq N$ with $A^+ \cap A^- = \emptyset$, a microbribery price function $\rho : N \times N \rightarrow \mathbb{N}$, and a positive integer $\ell$. \\
\textbf{Question:} &
Is there a way to obtain a profile $\varphi^\prime$ over $N$ by changing the qualifications in $\varphi$ for some pairs $M \subseteq N \times N$ of individuals such that $\rho(M) \leq \ell$ and $A^+ \subseteq f(N, \varphi^\prime)$ and $A^- \cap f(N, \varphi^\prime) = \emptyset$? \\
\hline
\end{tabularx}

In addition to the general $f$-\$GMB problem, there are the three special variants
\begin{itemize}
\item
$f$-\textsc{\$Constructive Group Microbribery} (\$CGMB) where the set $A^-$ is dropped,
\item
$f$-\textsc{\$Destructive Group Microbribery} (\$DGMB) where the set $A^+$ is dropped,
\item
$f$-\textsc{\$Exact Group Microbribery} (\$EGMB) where we add the premise $A^+ \cup A^- = N$.
\end{itemize}

Also, there are the four unpriced versions $f$-\textsc{Group Microbribery} (GMB), $f$-\textsc{Constructive Group Microbribery} (CGMB), $f$-\textsc{Destructive Group Microbribery} (DGMB), and $f$-\textsc{Exact Group Microbribery} (EGMB) where $\rho$ assigns price $1$ to every pair of individuals, i.e.\@ $\rho\big((a, b)\big) = 1$ for all $(a, b) \in N \times N$.

\subsection{Some remarks on naming, notation, definitions, and problem instances}

It should be noted that the above definitions and problem names differ slightly from what is found in other papers.

\textcite{YD18} only study the constructive group control problems. Therefore, they refer to $f$-\textsc{Constructive Group Control by Adding Individuals} (resp.\@ \textsc{Deleting Individuals} / \textsc{Partitioning of Individuals}) simply as $f$-\textsc{Group Control by Adding Individuals} (resp.\@ \textsc{Deleting Individuals} / \textsc{Partitioning of Individuals}). Also, they use $S$ instead of $A^+$ to denote the target set, and they require $S$ to be nonempty. In contrast, we allow $A^+$ to be empty as it makes some generalizations and reductions easier.

Similarly, \textcite{ERY20} require their target set $S$ to be nonempty in destructive group control and bribery, whereas we allow our target set $A^-$ to be empty.

\textcite{BBKL20} use $A$ instead of $N$ to denote the set of individuals. Moreover, they refer to the general problems as \textsc{Constructive+Destructive}; and they use the names \textsc{Agent Bribery} and \textsc{Link Bribery} instead of \textsc{Group Bribery} and \textsc{Group Microbribery}, respectively.

Note that for all NP-completeness results in this work, we only show NP-hardness. It is easy to verify that all considered problems are in NP because the socially qualified individuals for any set $N$ and profile $\varphi$ can always be calculated in polynomial time.

Furthermore, note that we only consider problem instances where at least one individual from $A^+$ is initially socially disqualified, i.e.\@ $A^+ \not\subseteq f(N, \varphi)$, and at least one individual from $A^-$ is initially socially qualified, i.e.\@ $A^- \cap f(N, \varphi) \neq \emptyset$. The only exception to this rule is when one of the sets $A^+$ or $A^-$ is empty (which we also allow).

For the general and exact problems, it could be interesting to also consider instances where both $A^+$ and $A^-$ are nonempty, but only one of them fulfills the above condition. For example, consider an instance where all individuals from $A^+$ are already socially qualified, and the attacker only needs to take care of the individuals from $A^-$ (without disqualifying someone from $A^+$ in the process). However, such instances are not considered in this work.

\clearpage

\subsection{Problem Acronyms Overview}

\begin{table}[!htb]
\begin{tabularx}{\textwidth}{lX}
\hline
Acronym & Problem Name \\
\hline

CGB &
(Unpriced) Constructive Group Bribery \\

CGCAI &
Constructive Group Control by Adding Individuals \\

CGCDI &
Constructive Group Control by Deleting Individuals \\

CGCPI &
Constructive Group Control by Partitioning of Individuals \\

CGMB &
(Unpriced) Constructive Group Microbribery \\

\hline

DGB &
(Unpriced) Destructive Group Bribery \\

DGCAI &
Destructive Group Control by Adding Individuals \\

DGCDI &
Destructive Group Control by Deleting Individuals \\

DGCPI &
Destructive Group Control by Partitioning of Individuals \\

DGMB &
(Unpriced) Destructive Group Microbribery \\

\hline

EGB &
(Unpriced) Exact Group Bribery \\

EGCAI &
Exact Group Control by Adding Individuals \\

EGCPI &
Exact Group Control by Partitioning of Individuals \\

EGMB &
(Unpriced) Exact Group Microbribery \\

\hline

GB &
(Unpriced) Group Bribery \\

GCAI &
Group Control by Adding Individuals \\

GCDI &
Group Control by Deleting Individuals \\

GCPI &
Group Control by Partitioning of Individuals \\

GMB &
(Unpriced) Group Microbribery \\

\hline

\$CGB &
(Priced) \$Constructive Group Bribery \\

\$CGMB &
(Priced) \$Constructive Group Microbribery \\

\$DGB &
(Priced) \$Destructive Group Bribery \\

\$DGMB &
(Priced) \$Destructive Group Microbribery \\

\$EGB &
(Priced) \$Exact Group Bribery \\

\$EGMB &
(Priced) \$Exact Group Microbribery \\

\$GB &
(Priced) \$Group Bribery \\

\$GMB &
(Priced) \$Group Microbribery \\

\hline
\end{tabularx}
\caption{
An overview of the problem names and their acronyms used in this work.
}
\label{tab:problem_acronyms}
\end{table}

% table formatting from now on
\renewcommand\cellalign{lt}
\renewcommand{\arraystretch}{1.8}

\clearpage

\section{Binary profiles}
\label{sec:binary}

In this section, we provide an overview of the known complexity results for group control and bribery problems in binary profiles. We consider all problems listed in \secref{sec:problem_definitions} under each of the social rules defined in \secref{sec:social_rules}.

\subsection{Constructive group control and bribery}
\label{sec:binary_constructive}

We begin with the constrictive problems. The three constructive group control problems were first studied by \textcite{YD18}. \textcite{ERY20} first studied the unpriced version of constructive group bribery, and \textcite{BBKL20} extended their results to the priced problem versions and to microbribery. Additionally, the unpriced microbribery problem was also studied by \textcite{EY20}.

\tableref{tab:binary_constructive} lists the known complexity results for constructive group control and bribery.

\begin{table}[!htb]
\begin{tabularx}{\textwidth}{
l
p{0.07\textwidth}
X
X
X
X
X
X
X
}
\hline & \multicolumn{6}{l}{Consent rules $f^{(s, t)}$} & $f^{\text{CSR}}$ & $f^{\text{LSR}}$ \\
\cline { 2 - 7 } & $s=1$ & & & $s \geq 2$ & & & & \\
\cline { 2 - 7 } & $t=1$ & $t=2$ & $t \geq 3$ & $t=1$ & $t=2$ & $t \geq 3$ & & \\
\hline

\makecell{CGCAI \\ ~} &
I &
I &
I &
\makecell{NP-c \\ (+)} &
\makecell{NP-c \\ (+)} &
\makecell{NP-c \\ (+)} &
NP-c &
NP-c \\

\makecell{CGCDI \\ ~} &
I &
P &
\makecell{NP-c \\ (+)} &
I &
P &
\makecell{NP-c \\ (+)} &
P &
I \\

\makecell{CGCPI \\ ~} &
I &
NP-c &
NP-c &
I &
NP-c &
NP-c &
? &
I \\

\makecell{CGB \\ ~} &
P &
\makecell{NP-c \\ (+)} &
\makecell{NP-c \\ (+)} &
\makecell{NP-c \\ (+)} &
\makecell{NP-c \\ (+)} &
\makecell{NP-c \\ (+)} &
P &
P \\

\makecell{\$CGB \\ ~} &
P &
\makecell{NP-c \\ (+)} &
\makecell{NP-c \\ (+)} &
\makecell{NP-c \\ (+)} &
\makecell{NP-c \\ (+)} &
\makecell{NP-c \\ (+)} &
P &
P \\

\makecell{CGMB \\ ~} &
P &
P &
P &
P &
P &
P &
\makecell{NP-c \\ (+)} &
\makecell{NP-c \\ (+)} \\

\makecell{\$CGMB \\ ~} &
P &
P &
P &
P &
P &
P &
\makecell{NP-c \\ (+)} &
\makecell{NP-c \\ (+)} \\

\hline
\end{tabularx}
\caption{
A summary of the complexity results for constructive group control and bribery.
In the table, ``P'' stands for ``polynomial-time solvable'', ``NP-c'' stands for ``NP-complete'', ``I'' stands for ``immune'', and ``?'' means that the complexity of the problem is open.
The symbol ``+'' below a NP-completeness result indicates that the problem is fixed-parameter tractable (FPT) with respect to $|A^+|$.
}
\label{tab:binary_constructive}
\end{table}

The results for CGCAI, CGCDI and CGCPI are all from \textcite{YD18}. In particular, they show that $f^{(s, t)}$-CGCAI and -CGCDI both are FPT with respect to $|A^+|$ \cite[Theorem 9]{YD18} and that $f^{(s, t)}$-CGCPI is NP-hard even if $|A^+| = 1$ \cite[Theorem 10]{YD18}. For the FPT result, they use ILP formulations and the algorithm by \textcite{L83}. Many FPT algorithms for various group control and bribery problems can be obtained by this approach (see also \thmref{thm:general_control_fpt_aa}).

The NP-completeness of $f^{(s, t)}$-CGB when $t \geq 2$ is shown by \textcite[Theorem 2]{ERY20} via a reduction from \textsc{Vertex Cover}. It trivially extends to \$CGB.
When $t=1$ and $s \geq 2$, the problem is still NP-complete; this follows from the reduction from \textsc{Independent Set} by \textcite[Theorem 5]{BBKL20}, and we additionally show an alternative proof in \secref{sec:cgb_npc}.
However, \textcite[Theorem 1]{ERY20} show that $f^{(s, t)}$-CGB is XP with respect to $s$ when $t=1$ and $s \geq 2$.
In \secref{sec:constructive_param}, we show that this XP-algorithm also extends to \$CGB with a few modifications. Additionally, we observe that there exists a trivial linear-time algorithm for $f^{(1, 1)}$-\$CGB. The fixed-parameter tractability of $f^{(s, t)}$-CGB and -\$CGB with respect to $|A^+|$ is shown by \textcite[Theorem 8]{BBKL20} using another ILP formulation and the algorithm by \textcite{L83}.

\textcite[Theorem 3]{ERY20} show the polynomial-time result for $f^{\text{CSR}} / f^{\text{LSR}}$-CGB. \textcite[Theorem 1 and Corollary 1]{BBKL20} show how to extend it to \$CGB; their algorithm is based on solving \textsc{Minimum Weighted Separator} in an auxiliary graph.

The polynomial-time result for $f^{(s, t)}$-CGMB is shown by \textcite[Theorem 1]{EY20}. \textcite[Observation 2]{BBKL20} show it also for \$CGMB.

The NP-completeness of $f^{\text{CSR}} / f^{\text{LSR}}$-CGMB is shown by \textcite[Theorem 3]{BBKL20} via a reduction from \textsc{Set Cover}, and additionally by \textcite[Theorems 3 and 2]{EY20}. The results trivially extend to \$CGMB. \textcite[Theorem 4]{BBKL20} show the fixed-parameter tractability of $f^{\text{CSR}} / f^{\text{LSR}}$-CGMB and -\$CGMB with respect to $|A^+|$ by reducing to \textsc{Weighted Directed Steiner Tree}.

\textcite[Theorem 1]{ERY20} claim that $f^{(s, t)}$-CGB is polynomial-time solvable if $t=1$. In fact, this is not correct (it only holds if we treat $s$ as a constant). Below, we show that $f^{(s, t)}$-CGB is actually NP-complete for all $s \geq 2$ and $t=1$.

\subsubsection{Constructive group bribery for consent rules with \texorpdfstring{$s \geq 2$}{} and \texorpdfstring{$t = 1$}{} is NP-complete}
\label{sec:cgb_npc}

\textcite[Theorem 5]{BBKL20} show a reduction from \textsc{Independent Set} to $f^{(s, t)}$-\textsc{Constructive Group Bribery}. The \textsc{Independent Set} problem asks whether a given graph contains a set of pairwise non-adjacent vertices of size $k$. In the reduced $f^{(s, t)}$-CGB instance, $t$ and $s$ are set to $t=1$ and $s=k+2$, respectively. From this and the fact that the \textsc{Independent Set} problem is NP-complete \cite{GJ79}, it immediately follows that both $f^{(s, t)}$-CGB and $f^{(s, t)}$-\$CGB are NP-complete even if $t=1$.

\begin{Theorem} \label{thm:cgb_npc}
$f^{(s, t)}$-\textsc{Constructive Group Bribery} and $f^{(s, t)}$-\textsc{\$Constructive Group Bribery} are NP-complete for all $s \geq 2$ and $t=1$.
\end{Theorem}

Below, we provide an alternative proof for \thmref{thm:cgb_npc} by reducing from \textsc{Restricted Exact Cover by 3-sets} (RX3C). With this reduction, we can show that the problem is still NP-complete even when each individual in $A^+$ is only missing a single qualification to become socially qualified and when the choices to provide this qualification are limited to just three (see \secref{sec:constructive_param}).

For the RX3C problem, we are given a set $X$ of size $3m$ and a family $\mathcal{F}$ of subsets of $X$ such that each subset $F \in \mathcal{F}$ contains exactly three elements from $X$ and each element in $X$ is contained in exactly three triplets $F \in \mathcal{F}$. The question is whether there is a subfamily $\mathcal{F}^\prime \subseteq \mathcal{F}$ such that every element in $X$ is contained in exactly one triplet in $\mathcal{F}^\prime$. This problem is known to be NP-complete \cite[Theorem A.1]{G85}. We now provide a proof for \thmref{thm:cgb_npc} by reducing from RX3C:

\begin{Proof}
Given a RX3C-instance $(X, \mathcal{F})$ with $|X|=3m$ (and thus $|\mathcal{F}|=3m$), we construct an instance of CGB as follows:

We introduce one individual $a_x$ for each element $x \in X$ and let $N_X = \{a_x : x \in X\}$. We also introduce one individual $a_F$ for each triplet $F \in \mathcal{F}$ and let $N_\mathcal{F} = \{a_F : F \in \mathcal{F}\}$. For all $x, x^\prime \in X$, we set $\varphi(a_x, a_{x^\prime}) = 1$. For all $F \in \mathcal{F}$ and $x \in X$, we set $\varphi(a_F, a_x) = 1$ if and only if $x \not\in F$. All other qualifications do not matter and can be set arbitrarily. Finally, we set $N = N_X \cup N_\mathcal{F}$, $A^+ = N_X$, $t=1$, $s = 6m - 2$, and $\ell=m$. Note that each individual $a_x \in N_X$ is qualified by exactly $3m + 3m - 3$ individuals (including themself) and is therefore missing exactly one qualification to become socially qualified.

We now show that there exists a successful bribery of cost at most $\ell$ in the constructed instance if and only if there exists an exact 3-set cover for $X$ in $\mathcal{F}$.

($\Rightarrow$) Assume that $\mathcal{F}^\prime \subseteq \mathcal{F}$ is an exact 3-set cover for $X$. Since $|X| = 3m$ and each element in $X$ is contained in exactly one triplet in $\mathcal{F}^\prime$, we know that $|\mathcal{F}^\prime| = m = \ell$. We bribe all individuals $a_F$ where $F \in \mathcal{F}^\prime$ to qualify everyone. Thereby, every individual $a_x \in N_X$ gains exactly one qualification (because $\mathcal{F}^\prime$ is an exact 3-set cover for $X$). Thus, all individuals in $A^+$ are now socially qualified.

($\Leftarrow$) Assume we are given a successful bribery $N^\prime \subseteq N$ consisting of $\ell = m$ individuals. We claim that no individual $a_x \in N_X$ can be part of $N^\prime$. Assume that some $a_x$ is part of $N^\prime$, then at most $m-1$ individuals from $N_\mathcal{F}$ can be bribed. Since $a_x$ already qualified all individuals in $N_X$ before the bribery, at most $3(m-1)$ individuals in $N_X$ can gain an additional qualification after the bribery, leaving at least three individuals in $A^+$ not socially qualified. Therefore, $N^\prime$ must consist of $m$ individuals from $N_\mathcal{F}$. From this it follows that $\mathcal{F}^\prime = \{F \in \mathcal{F} : a_F \in N^\prime\}$ is an exact 3-set cover for $X$.

Because we can reduce from CGB to \$CGB, the NP-completeness extends also to the priced version.
\end{Proof}

\subsubsection{Parameterized Complexity}
\label{sec:constructive_param}

We now turn to the parameterized complexity of constructive group control and bribery problems. Note that in the CGB-instances constructed in the proof for \thmref{thm:cgb_npc}, each individual in $A^+$ is only missing a single qualification to become socially qualified. Furthermore, for each individual in $A^+$ there are only three individuals who could be bribed to provide that additional qualification. Even in this restricted setting, the problem remains NP-complete. This observation motivates the formalization of a special parameter for instances of $f^{(s, t)}$-CGB and -\$CGB with $t=1$.

Let $N$ be a set of individuals and let $A^+ \subseteq N$. Let $\varphi$ be a profile over $N$ where $\varphi(a, a) = 1$ for all $a \in A^+$. For each $a \in A^+$ let $s^\text{missing}(a) = \operatorname{max}{(0, s - |N^1_\varphi(a)|)}$ denote the number of additional qualifications $a$ needs to get to become socially qualified. Furthermore, let $s^\text{choices}(a) = |N^{-1}_\varphi(a)|$ denote the number of individuals who could be bribed to provide additional qualifications for $a$. We now define the parameter $s^\ast$ as follows:
$$
s^\ast = \operatorname{max}_{a \in A^+}{\big( s^\text{choices}(a) - s^\text{missing}(a) \big)}
$$

The parameter $s^\ast$ denotes the maximum difference between the number of missing qualifications and the number of choices we have for any individual in $A^+$. In the CGB-instances constructed in the proof for \thmref{thm:cgb_npc}, we have $s^\text{missing}(a) = 1$ and $s^\text{choices}(a) = 3$ for all $a \in A^+$. Hence, we obtain the following result:

\begin{Corollary} \label{cor:cgb_param}
$f^{(s, t)}$-\textsc{Constructive Group Bribery} and $f^{(s, t)}$-\textsc{\$Constructive Group Bribery} are NP-complete for all $s \geq 2$ and $t=1$ even if $s^\ast = 2$.
\end{Corollary}

Even though $f^{(s, t)}$-\textsc{Constructive Group Bribery} is NP-complete for all $s \geq 2$ and $t=1$, there is a simple XP-algorithm for it that runs in time $\mathcal{O}(n^{s+2})$ \cite[Theorem 1]{ERY20}. Here, we show how to extend this result to the priced version of \textsc{Constructive Group Bribery}.

\begin{Theorem} \label{thm:cgb_xp}
If $t=1$, $f^{(s, t)}$-\textsc{\$Constructive Group Bribery} can be solved in time $\mathcal{O}(n^{s+2})$ where $n = |N|$ denotes the number of individuals.
\end{Theorem}

\begin{Proof}
Given the $f^{(s, t)}$-\$CGB instance $(N, \varphi, A^+, \rho, \ell)$, we first compute
$$
A^+_{-1} = \{a \in A^+ : \varphi(a, a) = -1\}.
$$
Since $t=1$, we must bribe all individuals in $A^+_{-1}$ to qualify themselves. Therefore, if $\rho(A^+_{-1}) > \ell$ we immediately conclude that the instance is a NO-instance. Otherwise, we bribe all individuals in $A^+_{-1}$ and make then qualify everyone, including themselves. We update $\ell := \ell - |A^+_{-1}|$.

Now we need to ensure that every individual in $A^+$ is qualified by at least $s$ individuals. Note that it is never necessary to bribe more than $s$ individuals to achieve this. Hence, we iterate over all subsets $U \subseteq N$ of size $|U| \leq s$. For each such subset, we obtain an altered profile $\varphi^\prime$ by bribing the individuals in $U$ to qualify everyone. We then check if $\rho(U) \leq \ell$ and $A^+ \subseteq f^{(s, t)}(N, \varphi^\prime)$. If we find a subset $U$ that satisfies these conditions, we know that the instance is a YES-instance. Otherwise, it must be a NO-instance.

The checks for any subset $U \subseteq N$ can be done in time $\mathcal{O}(n^2)$. As there are $\binom{n}{\leq s} \in \mathcal{O}(n^s)$ subsets to consider, we get a total running time of $\mathcal{O}(n^{s+2})$.
\end{Proof}

When $s=t=1$, the social qualification of an individual only depends on their opinion about themselves. Thus, $f^{(s, t)}$-\$CGB is linear-time solvable in this case:

\begin{Observation}
$f^{(s, t)}$-\textsc{\$Constructive Group Bribery} is linear-time solvable if $s=t=1$.
\end{Observation}

It is easy to see that for all problems with a budget parameter $\ell$, there exist trivial XP-algorithms with respect to $\ell$:

\begin{Proposition} \label{prop:constructive_xp_l}
$f$-\textsc{Constructive Group Control by Adding Individuals}, \\ $f$-\textsc{Constructive Group Control by Deleting Individuals}, $f$-\textsc{Constructive Group Bribery} and $f$-\textsc{\$Constructive Group Bribery} can be solved in time $\mathcal{O}(n^{\ell+2})$. \\
$f$-\textsc{Constructive Group Microbribery} and $f$-\textsc{\$Constructive Group Microbribery} can be solved in time $\mathcal{O}(n^{2\ell+2})$.
\end{Proposition}

Note that some of the above results had already been listed by \textcite[Figure 1]{BBKL20}, but there were no proofs provided for them. Below, we present the proofs for $f$-\textsc{\$Constructive Group Bribery} and $f$-\textsc{\$Constructive Group Microbribery}. The proofs for the other problems work similarly.

\begin{Proof} \textbf{($f$-\$CGB)} ~
Given an $f$-\$CGB instance $(N, \varphi, A^+, \rho, \ell)$ with $n = |N|$, we guess the subset of individuals we must bribe to make everyone $A^+$ socially qualified. To do that, we iterate over all subsets $U \subseteq N$ of size $|U| \leq \ell$.

For each subset $U \subseteq N$, we obtain an altered profile $\varphi^\prime$ by bribing the individuals in $U$ to qualify everyone. We then check if $\rho(U) \leq \ell$ and $A^+ \subseteq f(N, \varphi^\prime)$. If we find a subset $U$ that satisfies these conditions, the instance is a YES-instance. Otherwise, we know that bribing any group of up to $\ell$ individuals is not sufficient to make all individuals in $A^+$ socially qualified. Since $\rho$ assigns a positive integer price to each individual, it follows that bribing any group of individuals with cost at most $\ell$ is not sufficient to make all individuals in $A^+$ socially qualified. Therefore, the instance must be a NO-instance.

The checks for any subset $U \subseteq N$ can be done in time $\mathcal{O}(n^2)$. As there are $\binom{n}{\leq \ell} \in \mathcal{O}(n^\ell)$ subsets to consider, we get a total running time of $\mathcal{O}(n^{\ell+2})$.
\end{Proof}

\begin{Proof} \textbf{($f$-\$CGMB)} ~
Given an $f$-\$CGMB instance $(N, \varphi, A^+, \rho, \ell)$ with $n = |N|$, we guess the subset of pairs of individuals whose valuations we need to change. To do that, we iterate over all subsets $M \subseteq N \times N$ of size $|M| \leq \ell$.

For each subset $M \subseteq N \times N$, we obtain an altered profile $\varphi^\prime$ by going over all pairs $(a, b) \in M$ and bribing each $a$ to qualify $b$. We then check if $\rho(M) \leq \ell$ and $A^+ \subseteq f(N, \varphi^\prime)$. If we find a subset $M$ that satisfies these conditions, the instance is a YES-instance. Otherwise, we know that changing the valuations for any set of up to $\ell$ pairs is not sufficient to make all individuals in $A^+$ socially qualified. Since $\rho$ assigns a positive integer price to each pair, it follows that changing the valuations for any set of pairs with cost at most $\ell$ is not sufficient to make all individuals in $A^+$ socially qualified. Therefore, the instance must be a NO-instance.

The checks for any subset $M \subseteq N \times N$ can be done in time $\mathcal{O}(n^2)$. As there are $\binom{n^2}{\leq \ell} \in \mathcal{O}(n^{2\ell})$ subsets to consider, we get a total running time of $\mathcal{O}(n^{2\ell+2})$.
\end{Proof}

In addition to being XP with respect to $\ell$ (as shown above), $f^{\text{CSR}} / f^{\text{LSR}}$-CGMB and -\$CGMB are also W[2]-hard with respect to $\ell$ \cite[Theorem 3]{BBKL20}.

For $f^{(s, t)}$-CGB and -\$CGB, we can also parameterize by $s$ and $t$, in addition to $\ell$. \figref{fig:spiderweb_constructive} provides an overview of the parameterized complexity of CGB and \$CGB for the consent rules with different parameter combinations.

In summary, $f^{(s, t)}$-CGB and -\$CGB are

\begin{itemize}[nolistsep]
\item
para-NP-hard with respect to $s+t$
(\tikz\draw [densely dotted,color=darkgray,line width=1.5pt,opacity=0.6] (0,0) -- (0.4,0);, left).
This follows from the NP-hardness shown by \textcite[Theorem 2]{ERY20} for all $s \geq 1$ and $t \geq 2$.

\item
W[2]-hard with respect to $\ell$ even when $s=1$
(\tikz\draw [color=red,line width=1.5pt,opacity=0.6] (0,0) -- (0.4,0);, left) \cite[Theorem 6]{BBKL20}.

\item
W[1]-hard with respect to $\ell+s$ even when $t=1$
(\tikz\draw [color=orange,line width=1.5pt,opacity=0.5] (0,0) -- (0.4,0);, left) \cite[Theorem 5]{BBKL20}.

\item
XP with respect to $s$ if $t=1$
(\tikz\draw [color=blue,line width=1.5pt,opacity=0.6] (0,0) -- (0.4,0);, right).
We show this in \thmref{thm:cgb_xp}.

\item
XP with respect to $\ell$
(\tikz\draw [densely dotted,color=blue,line width=1.5pt,opacity=0.6] (0,0) -- (0.4,0);, right).
We show this in \propref{prop:constructive_xp_l}.

\item
FPT with respect to $\ell+t$ when $s$ is a constant
(\tikz\draw [color=green,line width=1.5pt,opacity=0.6] (0,0) -- (0.4,0);, right) \cite[Theorem 7]{BBKL20}.
\end{itemize}

\clearpage

\newcommand{\D}{3} % number of dimensions (config option)
\newcommand{\U}{4} % number of scale units (config option)

\newdimen\R % maximal diagram radius (config option)
\R=3.5cm 
\newdimen\L % radius to put dimension labels (config option)
\L=4cm

\newcommand{\A}{360/\D} % calculated angle between dimension axes  

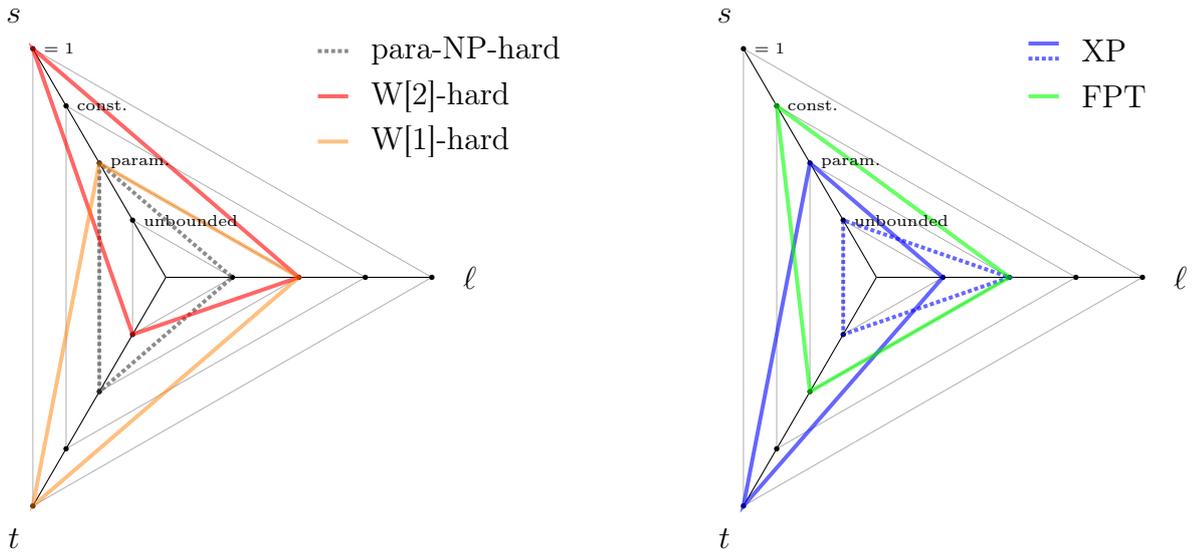
\begin{figure}[!htb]
\centering
\begin{minipage}[b]{.45\textwidth}

\begin{tikzpicture}[scale=1] %%%%%%%%%%%%%%%%%%%%%%%%%%%%%%%%%%%%%%%%%%

\spiderweb

% para-NP-hard with respect to s+t
\draw [densely dotted,color=darkgray,line width=1.5pt,opacity=0.6]
    (D1-2) --
    (D2-2) --
    (D3-1) -- cycle;

% W[2]-hard with respect to l even when s=1
\draw [color=red,line width=1.5pt,opacity=0.6]
    (D1-4) --
    (D2-1) --
    (D3-2) -- cycle;

% W[1]-hard with respect to l+s even when t=1
\draw [color=orange,line width=1.5pt,opacity=0.5]
    (D1-2) --
    (D2-4) --
    (D3-2) -- cycle;

% labels for spiderweb
\spiderweblabels

% labels for each dimension axis
\path (1*\A:\L) node (L1) {$s$};
\path (2*\A:\L) node (L2) {$t$};
\path (3*\A:\L) node (L3) {$\ell$};

% color legend
\draw [densely dotted,color=darkgray,line width=1.5pt,opacity=0.6] (2,3) -- (2.4,3)
    node[right,black,opacity=1] {~para-NP-hard};
\draw [color=red,line width=1.5pt,opacity=0.6] (2,2.4) -- (2.4,2.4)
    node[right,black,opacity=1] {~W[2]-hard};
\draw [color=orange,line width=1.5pt,opacity=0.5] (2,1.8) -- (2.4,1.8)
    node[right,black,opacity=1] {~W[1]-hard};

\end{tikzpicture} %%%%%%%%%%%%%%%%%%%%%%%%%%%%%%%%%%%%%%%%%%

\end{minipage}\hfill
\begin{minipage}[b]{.45\textwidth}

\begin{tikzpicture}[scale=1] %%%%%%%%%%%%%%%%%%%%%%%%%%%%%%%%%%%%%%%%%%

\spiderweb

% XP with respect to s if t=1
\draw [color=blue,line width=1.5pt,opacity=0.6]
    (D1-2) --
    (D2-4) --
    (D3-1) -- cycle;

% XP with respect to l
\draw [densely dotted,color=blue,line width=1.5pt,opacity=0.6]
    (D1-1) --
    (D2-1) --
    (D3-2) -- cycle;

% FPT with respect to l+t when s is a constant
\draw [color=green,line width=1.5pt,opacity=0.6]
    (D1-3) --
    (D2-2) --
    (D3-2) -- cycle;

% labels for spiderweb
\spiderweblabels

% labels for each dimension axis
\path (1*\A:\L) node (L1) {$s$};
\path (2*\A:\L) node (L2) {$t$};
\path (3*\A:\L) node (L3) {$\ell$};

% color legend
\draw [color=blue,line width=1.5pt,opacity=0.6] (2,3.1) -- (2.4,3.1);
\draw [densely dotted,color=blue,line width=1.5pt,opacity=0.6] (2,2.9) -- (2.4,2.9)
    node[yshift=0.25em,right,black,opacity=1] {~XP};
\draw [color=green,line width=1.5pt,opacity=0.6] (2,2.4) -- (2.4,2.4)
    node[right,black,opacity=1] {~FPT};

\end{tikzpicture} %%%%%%%%%%%%%%%%%%%%%%%%%%%%%%%%%%%%%%%%%%

\end{minipage}

\caption{
Parameterized complexity of $f^{(s, t)}$-CGB and -\$CGB for different parameter combinations. The figure on the left shows hardness results and the figure on the right shows results for XP (slice-wise polynomial) and FPT (fixed-parameter tractable) algorithms.
}
\label{fig:spiderweb_constructive}
\end{figure}

\clearpage

\subsection{Destructive group control and bribery}
\label{sec:binary_destructive}

We now consider the destructive problems. The three destructive group control problems and the unpriced bribery problem were first studied by \textcite{ERY20}. \textcite{BBKL20} then extended their results to the priced problem versions and to microbribery.

\tableref{tab:binary_destructive} lists the known complexity results for destructive group control and bribery.

\begin{table}[!htb]
\begin{tabularx}{\textwidth}{
l
p{0.07\textwidth}
X
X
X
X
X
X
X
}
\hline & \multicolumn{6}{l}{Consent rules $f^{(s, t)}$} & $f^{\text{CSR}}$ & $f^{\text{LSR}}$ \\
\cline { 2 - 7 } & $s=1$ & & $s=2$ & & $s \geq 3$ & & & \\
\cline { 2 - 7 } & $t=1$ & $t \geq 2$ & $t=1$ & $t \geq 2$ & $t=1$ & $t \geq 2$ & & \\
\hline

\makecell{DGCAI \\ ~} &
I &
\makecell{NP-c \\ ($\mathtt{\sim}$)} &
I &
\makecell{NP-c \\ ($\mathtt{\sim}$)} &
I &
\makecell{NP-c \\ ($\mathtt{\sim}$)} &
NP-c &
I \\

\makecell{DGCDI \\ ~} &
I &
I &
P &
P &
\makecell{NP-c \\ ($\mathtt{\sim}$)} &
\makecell{NP-c \\ ($\mathtt{\sim}$)} &
P &
P \\

\makecell{DGCPI \\ ~} &
I &
I &
NP-c &
NP-c &
NP-c &
NP-c &
? &
P \\

\makecell{DGB \\ ~} &
P &
\makecell{NP-c \\ ($\mathtt{\sim}$)} &
\makecell{NP-c \\ ($\mathtt{\sim}$)} &
\makecell{NP-c \\ ($\mathtt{\sim}$)} &
\makecell{NP-c \\ ($\mathtt{\sim}$)} &
\makecell{NP-c \\ ($\mathtt{\sim}$)} &
P &
P \\

\makecell{\$DGB \\ ~} &
P &
\makecell{NP-c \\ ($\mathtt{\sim}$)} &
\makecell{NP-c \\ ($\mathtt{\sim}$)} &
\makecell{NP-c \\ ($\mathtt{\sim}$)} &
\makecell{NP-c \\ ($\mathtt{\sim}$)} &
\makecell{NP-c \\ ($\mathtt{\sim}$)} &
P &
P \\

\makecell{DGMB \\ ~} &
P &
P &
P &
P &
P &
P &
P &
P \\

\makecell{\$DGMB \\ ~} &
P &
P &
P &
P &
P &
P &
P &
P \\

\hline
\end{tabularx}
\caption{
A summary of the complexity results for destructive group control and bribery.
In the table, ``P'' stands for ``polynomial-time solvable'', ``NP-c'' stands for ``NP-complete'', ``I'' stands for ``immune'', and ``?'' means that the complexity of the problem is open.
The symbol ``$\mathtt{\sim}$'' below a NP-completeness result indicates that the problem is fixed-parameter tractable (FPT) with respect to $|A^-|$.
}
\label{tab:binary_destructive}
\end{table}

The results for the consent rules $f^{(s, t)}$ can be derived using \textcite[Lemma 1]{ERY20}. Their Lemma basically states that constructive group control and bribery problems can be reduced to destructive group control and bribery problems by reversing the values in $\varphi$ and swapping $s$ and $t$:

\begin{Lemma} \cite{ERY20} ~
$f^{(s, t)}(N, \varphi) = N \setminus f^{(t, s)}(N, -\varphi)$ where $-\varphi$ is obtained from $\varphi$ by reversing the values (i.e.\@, $\varphi(a, b) = 1$ if and only if $-\varphi(a, b) = -1$).
\label{lemma:one}
\end{Lemma}

Regarding the results for $f^{(s, t)}$-DGCAI, $f^{(s, t)}$-DGCDI, $f^{(s, t)}$-DGCPI, and $f^{(s, t)}$-DGB, we refer to \textcite[Theorem 4]{ERY20}.
The results for $f^{(s, t)}$-\$DGB follow from \lemmaref{lemma:one} in a similar manner.
The polynomial-time solvability of $f^{(s, t)}$-DGMB and -\$DGMB follows from \textcite[Observation 2]{BBKL20}.
The fixed-parameter tractability of $f^{(s, t)}$-DGB and -\$DGB with respect to $|A^-|$ follows from \textcite[Corollary 2]{BBKL20}.

The results for $f^{\text{CSR}} / f^{\text{LSR}}$-DGCAI are shown by \textcite[Theorems 5 and 6]{ERY20}. For the NP-hardness result, they reduce from \textsc{Restricted Exact Cover by 3-sets}.

\textcite[Theorem 7]{ERY20} show the polynomial-time result for $f^{\text{CSR}} / f^{\text{LSR}}$-DGCDI by giving a reduction to the \textsc{Minimum $(u,v)$-separator} problem.

\textcite[Theorem 8]{ERY20} show a linear-time algorithm for solving $f^{\text{LSR}}$-DGCPI. The complexity of $f^{\text{CSR}}$-DGCPI is open, but \textcite[Theorem 9]{ERY20} show it is polynomial-time solvable when the profile $\varphi$ is symmetric.

The polynomial-time results for $f^{\text{CSR}} / f^{\text{LSR}}$-DGB are shown by \textcite[Theorem 10]{ERY20}. \textcite[Theorem 1 and Corollary 1]{BBKL20} show that they extend to \$DGB.

The polynomial-time solvability of $f^{\text{CSR}} / f^{\text{LSR}}$-DGMB and -\$DGMB is shown by \textcite[Theorem 2]{BBKL20}. Their algorithm works by solving the \textsc{Minimum Weighted Cut} problem in an auxiliary graph.

\subsubsection{Parameterized Complexity}
\label{sec:destructive_param}

We now turn to the parameterized complexity of destructive group control and bribery problems. Analogous to the constructive case, we can define a parameter $t^\ast$ for instances of $f^{(s, t)}$-DGB and -\$DGB with $s=1$. This parameter serves as a measure of how many disqualifications the individuals are missing and how many choices we have to provide them. Let $N$ be a set of individuals and let $A^- \subseteq N$. Let $\varphi$ be a profile over $N$ where $\varphi(a, a) = -1$ for all $a \in A^-$. For each $a \in A^-$ let $t^\text{missing}(a) = \operatorname{max}{(0, t - |N^{-1}_\varphi(a)|)}$ denote the number of additional disqualifications $a$ needs to get to become socially disqualified. Furthermore, let $t^\text{choices}(a) = |N^1_\varphi(a)|$ denote the number of individuals who could be bribed to provide additional disqualifications for $a$. We now define the parameter $t^\ast$ as follows:
$$
t^\ast = \operatorname{max}_{a \in A^-}{\big( t^\text{choices}(a) - t^\text{missing}(a) \big)}
$$

From \corref{cor:cgb_param} and \lemmaref{lemma:one}, we obtain the following result:

\begin{Corollary} \label{cor:dgb_param}
$f^{(s, t)}$-\textsc{Destructive Group Bribery} and $f^{(s, t)}$-\textsc{\$Destructive Group Bribery} are NP-complete for all $t \geq 2$ and $s=1$ even if $t^\ast = 2$.
\end{Corollary}

Also analogous to the constructive case (see \propref{prop:constructive_xp_l}), for all problems with a budget parameter $\ell$, there exist trivial XP-algorithms with respect to $\ell$:

\begin{Proposition} \label{prop:destructive_xp_l}
$f$-\textsc{Destructive Group Control by Adding Individuals}, \\ $f$-\textsc{Destructive Group Control by Deleting Individuals}, $f$-\textsc{Destructive Group Bribery} and $f$-\textsc{\$Destructive Group Bribery} can be solved in time $\mathcal{O}(n^{\ell+2})$. \\
$f$-\textsc{Destructive Group Microbribery} and $f$-\textsc{\$Destructive Group Microbribery} can be solved in time $\mathcal{O}(n^{2\ell+2})$.
\end{Proposition}

\begin{figure}[!htb]
\centering
\begin{minipage}[b]{.45\textwidth}

\begin{tikzpicture}[scale=1] %%%%%%%%%%%%%%%%%%%%%%%%%%%%%%%%%%%%%%%%%%

\spiderweb

% para-NP-hard with respect to s+t
\draw [densely dotted,color=darkgray,line width=1.5pt,opacity=0.6]
    (D1-2) --
    (D2-2) --
    (D3-1) -- cycle;

% W[2]-hard with respect to l even when t=1
\draw [color=red,line width=1.5pt,opacity=0.6]
    (D1-1) --
    (D2-4) --
    (D3-2) -- cycle;

% W[1]-hard with respect to l+t even when s=1
\draw [color=orange,line width=1.5pt,opacity=0.5]
    (D1-4) --
    (D2-2) --
    (D3-2) -- cycle;

% labels for spiderweb
\spiderweblabels

% labels for each dimension axis
\path (1*\A:\L) node (L1) {$s$};
\path (2*\A:\L) node (L2) {$t$};
\path (3*\A:\L) node (L3) {$\ell$};

% color legend
\draw [densely dotted,color=darkgray,line width=1.5pt,opacity=0.6] (2,3) -- (2.4,3)
    node[right,black,opacity=1] {~para-NP-hard};
\draw [color=red,line width=1.5pt,opacity=0.6] (2,2.4) -- (2.4,2.4)
    node[right,black,opacity=1] {~W[2]-hard};
\draw [color=orange,line width=1.5pt,opacity=0.5] (2,1.8) -- (2.4,1.8)
    node[right,black,opacity=1] {~W[1]-hard};

\end{tikzpicture} %%%%%%%%%%%%%%%%%%%%%%%%%%%%%%%%%%%%%%%%%%

\end{minipage}\hfill
\begin{minipage}[b]{.45\textwidth}

\begin{tikzpicture}[scale=1] %%%%%%%%%%%%%%%%%%%%%%%%%%%%%%%%%%%%%%%%%%

\spiderweb

% XP with respect to t if s=s
\draw [color=blue,line width=1.5pt,opacity=0.6]
    (D1-4) --
    (D2-2) --
    (D3-1) -- cycle;

% XP with respect to l
\draw [densely dotted,color=blue,line width=1.5pt,opacity=0.6]
    (D1-1) --
    (D2-1) --
    (D3-2) -- cycle;

% FPT with respect to l+s when t is a constant
\draw [color=green,line width=1.5pt,opacity=0.6]
    (D1-2) --
    (D2-3) --
    (D3-2) -- cycle;

% labels for spiderweb
\spiderweblabels

% labels for each dimension axis
\path (1*\A:\L) node (L1) {$s$};
\path (2*\A:\L) node (L2) {$t$};
\path (3*\A:\L) node (L3) {$\ell$};

% color legend
\draw [color=blue,line width=1.5pt,opacity=0.6] (2,3.1) -- (2.4,3.1);
\draw [densely dotted,color=blue,line width=1.5pt,opacity=0.6] (2,2.9) -- (2.4,2.9)
    node[yshift=0.25em,right,black,opacity=1] {~XP};
\draw [color=green,line width=1.5pt,opacity=0.6] (2,2.4) -- (2.4,2.4)
    node[right,black,opacity=1] {~FPT};

\end{tikzpicture} %%%%%%%%%%%%%%%%%%%%%%%%%%%%%%%%%%%%%%%%%%

\end{minipage}

\caption{
Parameterized complexity of $f^{(s, t)}$-DGB and -\$DGB for different parameter combinations. The figure on the left shows hardness results and the figure on the right shows results for XP (slice-wise polynomial) and FPT (fixed-parameter tractable) algorithms.
}
\label{fig:spiderweb_destructive}
\end{figure}
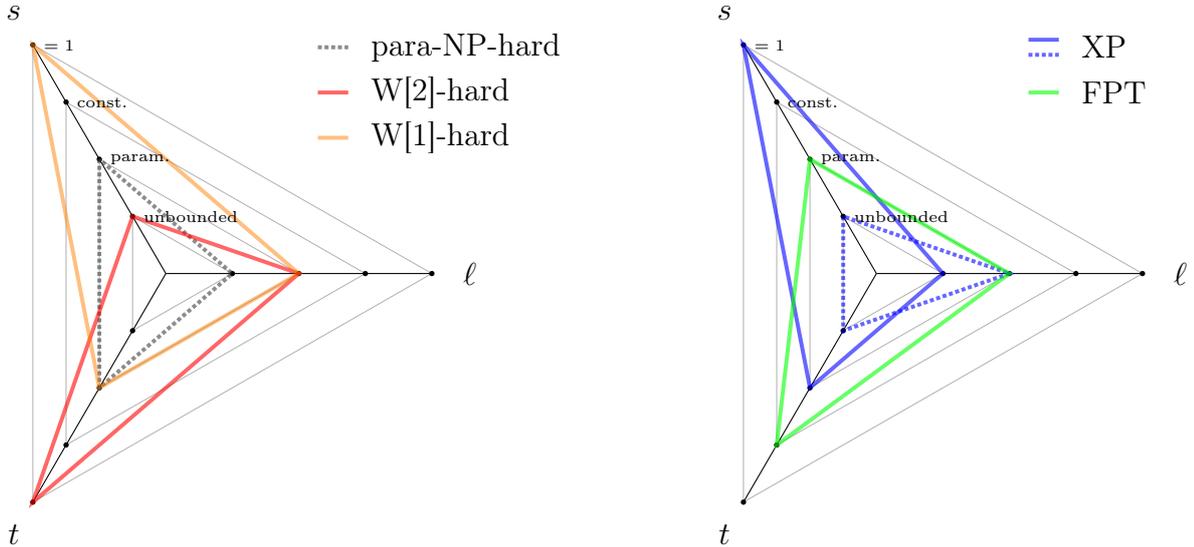

For $f^{(s, t)}$-DGB and -\$DGB, we can also parameterize by $s$ and $t$, in addition to $\ell$. \figref{fig:spiderweb_destructive} provides an overview of the parameterized complexity of DGB and \$DGB for the consent rules with different parameter combinations.

In summary, $f^{(s, t)}$-DGB and -\$DGB are

\begin{itemize}[nolistsep]
\item
para-NP-hard with respect to $s+t$
(\tikz\draw [densely dotted,color=darkgray,line width=1.5pt,opacity=0.6] (0,0) -- (0.4,0);, left).
This follows from the NP-hardness shown by \textcite[Theorem 4]{ERY20} for all $s \geq 2$ and $t \geq 1$.

\item
W[2]-hard with respect to $\ell$ even when $t=1$
(\tikz\draw [color=red,line width=1.5pt,opacity=0.6] (0,0) -- (0.4,0);, left) \cite[Corollary 2]{BBKL20}.

\item
W[1]-hard with respect to $\ell+t$ even when $s=1$
(\tikz\draw [color=orange,line width=1.5pt,opacity=0.5] (0,0) -- (0.4,0);, left) \cite[Corollary 2]{BBKL20}.

\item
XP with respect to $t$ if $s=1$
(\tikz\draw [color=blue,line width=1.5pt,opacity=0.6] (0,0) -- (0.4,0);, right).
Follows from \thmref{thm:cgb_xp} and \lemmaref{lemma:one}.

\item
XP with respect to $\ell$
(\tikz\draw [densely dotted,color=blue,line width=1.5pt,opacity=0.6] (0,0) -- (0.4,0);, right).
See \propref{prop:destructive_xp_l}.

\item
FPT with respect to $\ell+s$ when $t$ is a constant
(\tikz\draw [color=green,line width=1.5pt,opacity=0.6] (0,0) -- (0.4,0);, right) \cite[Corollary 2]{BBKL20}.
\end{itemize}

\clearpage

\subsection{Exact group control and bribery}
\label{sec:binary_exact}

We now consider manipulative attacks where the attacker has an exact objective, i.e.\@ they can precisely specify what the final subgroup of socially qualified individuals should be.

The exact group control problems EGCAI and EGCPI have not been studied up to now. However, in \secref{sec:exact_observations}, we present some easily obtainable results that cover most of the cases. The bribery and microbribery problems were first studied by \textcite{BBKL20}.

\tableref{tab:binary_exact} lists the known complexity results for exact group control and bribery.

\begin{table}[!htb]
\begin{tabularx}{\textwidth}{
p{0.12\textwidth}
p{0.1\textwidth}
X
X
p{0.1\textwidth}
p{0.1\textwidth}
X
}
\hline & \multicolumn{4}{l}{Consent rules $f^{(s, t)}$} & $f^{\text{CSR}}$ & $f^{\text{LSR}}$ \\
\cline { 2 - 5 } & $s=1$ & & $s \geq 2$ & & & \\
\cline { 2 - 5 } & $t=1$ & $t \geq 2$ & $t=1$ & $t \geq 2$ & & \\
\hline

\makecell{EGCAI \\ ~} &
I &
\makecell{I if $A^+ \neq \emptyset$ \\ NP-c else} &
\makecell{I if $A^- \neq \emptyset$ \\ NP-c else} &
NP-c &
NP-c &
\makecell{I if $A^- \neq \emptyset$ \\ NP-c else} \\

\makecell{EGCPI \\ ~} &
I &
I &
\makecell{I if $A^+ \neq \emptyset$ \\ ? else} &
NP-c &
? &
I \\

\makecell{EGB \\ ~} &
P &
NP-c &
NP-c &
NP-c &
P &
P \\

\makecell{\$EGB \\ ~} &
P &
NP-c &
NP-c &
NP-c &
P &
P \\

\makecell{EGMB \\ ~} &
P &
P &
P &
P &
P &
P \\

\makecell{\$EGMB \\ ~} &
P &
P &
P &
P &
P &
P \\

\hline
\end{tabularx}
\caption{
A summary of the complexity results for exact group control and bribery.
In the table, ``P'' stands for ``polynomial-time solvable'', ``NP-c'' stands for ``NP-complete'', ``I'' stands for ``immune'', and ``?'' means that the complexity of the problem is open.
}
\label{tab:binary_exact}
\end{table}

The results for $f^{(s, t)}$-EGB and -\$EGB are shown by \textcite[Observation 3]{BBKL20}. In particular, the NP-completeness for all $s \geq 1$ and $t \geq 2$ follows from the NP-completeness proof for $f^{(s, t)}$-CGB by \textcite[Theorem 2]{ERY20} where they set $A^+ = N$ (i.e.\@ all individuals must be socially qualified). Similarly, the NP-completeness for all $s \geq 2$ and $t \geq 1$ follows from the same proof for $f^{(s, t)}$-DGB in conjunction with \lemmaref{lemma:one}.

\textcite[Theorem 1 and Corollary 1]{BBKL20} show the polynomial-time results for $f^{\text{CSR}} / f^{\text{LSR}}$-EGB and -\$EGB via a reduction to \textsc{Minimum Weighted Separator}.

The polynomial-time solvability of $f^{(s, t)}$-EGMB and -\$EGMB follows from \textcite[Observation 2]{BBKL20}. For $f^{\text{CSR}} / f^{\text{LSR}}$-EGMB and -\$EGMB, the polynomial-time solvability is shown by \textcite[Theorem 2]{BBKL20} via a reduction to \textsc{Minimum Weighted Spanning Arborescence}.

In the following section, we present some easily obtainable results for the previously unstudied EGCAI and EGCPI problems.

\subsubsection{Complexity of exact group control problems}
\label{sec:exact_observations}

We first consider the consent rules $f^{(s, t)}$. When either of $s$ or $t$ is set to $1$, the consent rule is immune to most instances of EGCAI:

\begin{Theorem} \label{thm:fst_immune_egcai}
$f^{(s, t)}$ is immune to \textsc{Exact Group Control by Adding Individuals} when $s=1$ and $A^+ \neq \emptyset$, and also when $t=1$ and $A^- \neq \emptyset$.
\end{Theorem}

\renewcommand{\arraystretch}{1.2}

\begin{Proof}
The proof uses similar arguments as the proof for $f^{(s, t)}$-CGCAI by \textcite[Theorem 2]{YD18}. Our proof is based on two simple facts:

\textit{Fact \ref*{thm:fst_immune_egcai}.1}: When $s=1$, it is impossible to turn an individual socially qualified by adding individuals. To see this, let $a \in A^+$ be an individual who is not socially qualified initially, i.e.\@ $a \not\in f^{(s, t)}(T, \varphi)$. Because $s=1$, it must hold that $\varphi(a, a) = -1$ and there are already at least $t$ individuals in $T$ who disqualify $a$. Therefore, no matter which individuals we add to $T$, we can never make the individual $a$ socially qualified.

\textit{Fact \ref*{thm:fst_immune_egcai}.2}: When $t=1$, it is impossible to turn an individual socially disqualified by adding individuals. To see this, let $a \in A^-$ be an individual who is socially qualified, i.e.\@ $a \in f^{(s, t)}(T, \varphi)$. Because $t=1$, it must hold that $\varphi(a, a) = 1$ and there are already at least $s$ individuals in $T$ who qualify $a$. Therefore, no matter which individuals we add to $T$, we can never make the individual $a$ socially disqualified.
\end{Proof}

For all remaining cases not covered above, $f^{(s, t)}$-EGCAI is NP-complete. This follows from the fact that $f^{(s, t)}$-CGCAI is NP-complete for all $s \geq 2$ and $t \geq 1$ \cite[Theorem 4]{YD18}, and that $f^{(s, t)}$-DGCAI is NP-complete for all $s \geq 1$ and $t \geq 2$ \cite[Theorem 4]{ERY20}. In the corresponding proof by \textcite{YD18}, they set $A^+ = T$ (i.e.\@ all individuals must be socially qualified). In conjunction with \lemmaref{lemma:one}, it follows that the NP-completeness also applies when $A^- = T$ (i.e.\@ all individuals must be socially disqualified).

\begin{Corollary}
$f^{(s, t)}$-\textsc{Exact Group Control by Adding Individuals} is NP-complete for all $s \geq 2$ and $t \geq 2$. When $A^+ = \emptyset$, it is also NP-complete for $s=1$ and all $t \geq 2$; and when $A^- = \emptyset$, it is also NP-complete for $t=1$ and all $s \geq 2$.
\end{Corollary}

For EGCPI, we can obtain similar results as for EGCAI. When either of $s$ or $t$ is set to $1$, the consent rule is immune to most instances of EGCPI:

\begin{Theorem} \label{thm:fst_immune_egcpi}
$f^{(s, t)}$ is immune to \textsc{Exact Group Control by Partitioning of Individuals} when $s=1$ and $A^- \neq \emptyset$, and also when $t=1$ and $A^+ \neq \emptyset$.
\end{Theorem}

\begin{Proof}
The proof uses similar arguments as the proof for $f^{(s, t)}$-CGCPI by \textcite[Theorem 2]{YD18}. Again, our proof is based on two simple facts:

\textit{Fact \ref*{thm:fst_immune_egcpi}.1}: When $s=1$, it is impossible to turn an individual socially disqualified by partitioning (or deleting) individuals. To see this, let $a \in A^-$ be an individual who is socially qualified, i.e.\@ $a \in f^{(s, t)}(N, \varphi)$. We distinguish between two cases. \\
\begin{tabularx}{\textwidth}{lX}
Case 1.1: $\varphi(a, a) = 1$ &
Because $s=1$, it holds that $a \in f^{(s, t)}(U, \varphi)$ for all $U \subseteq N$ with $a \in U$.
\\
Case 1.2: $\varphi(a, a) = -1$ &
In this case, there are less than $t$ individuals in $N$ who disqualify $a$. Therefore, no matter how we partition the set $N$ (or which individuals we delete from $N$), we can never make the individual $a$ socially disqualified.
\\
\end{tabularx}

\textit{Fact \ref*{thm:fst_immune_egcpi}.2}: When $t=1$, it is impossible to turn an individual socially qualified by partitioning (or deleting) individuals. To see this, let $a \in A^+$ be an individual who is not socially qualified, i.e.\@ $a \not\in f^{(s, t)}(N, \varphi)$. We distinguish between two cases. \\
\begin{tabularx}{\textwidth}{lX}
Case 2.1: $\varphi(a, a) = 1$ &
In this case, there are less than $s$ individuals in $N$ who qualify $a$. Therefore, no matter how we partition the set $N$ (or which individuals we delete from $N$), we can never make the individual $a$ socially qualified.
\\
Case 2.2: $\varphi(a, a) = -1$ &
Because $t=1$, it holds that $a \not\in f^{(s, t)}(U, \varphi)$ for all $U \subseteq N$ with $a \in U$.
\\
\end{tabularx}
\end{Proof}

When $A^- = \emptyset$, all consent rules are immune to EGCPI:

\begin{Observation} \label{obs:fst_immune_egcpi}
For all combinations of $s$ and $t$, $f^{(s, t)}$ is immune to \textsc{Exact Group Control by Partitioning of Individuals} when $A^- = \emptyset$.
\end{Observation}

\begin{Proof}
When $A^- = \emptyset$, we have $A^+ = N$, i.e.\@ the task is to make everyone socially qualified. We fix any individual $a \in A^+$ who is initially socially disqualified, i.e.\@ $a \not\in f^{(s, t)}(N, \varphi)$. Because $A^+ = N$, all individuals must survive the first stage of the selection. But then in the second stage of the selection, the individual $a$ would always be socially disqualified.

Clearly, the above observation holds for all combinations of $s$ and $t$. Hence, all $f^{(s, t)}$ rules are immune to EGCPI when $A^- = \emptyset$.
\end{Proof}

The above results imply that all $f^{(s, t)}$ rules with $s=1$ are immune to EGCPI not only when $A^- \neq \emptyset$ (by \thmref{thm:fst_immune_egcpi}) but even when $A^- = \emptyset$ (by \obsref{obs:fst_immune_egcpi}).

\begin{Corollary} \label{cor:fst_immune_egcpi}
For $s=1$ and all $t \geq 1$, the $f^{(s, t)}$ rule is immune to EGCPI.
\end{Corollary}

However, when both $s \geq 2$ and $t \geq 2$, $f^{(s, t)}$-ECGPI is NP-complete:

\begin{Theorem} \label{thm:fst_egcpi_npc}
$f^{(s, t)}$-\textsc{Exact Group Control by Partitioning of Individuals} is NP-complete for all $s \geq 2$ and $t \geq 2$.
\end{Theorem}

\begin{Proof}
For all $s \geq 1$ and $t \geq 2$, \textcite[Theorem 5]{YD18} show a reduction from \textsc{3-SAT} to $f^{(s, t)}$-CGCPI. All resulting CGCPI instances where $s \geq 2$ can easily be turned into EGCPI instances:

In the reduction by \textcite{YD18}, they create individuals $a(x,1)$ and $a(x,2)$ for each variable $x \in X$, an individual $a(c)$ for each clause $c \in C$, and one individual $a(C)$. For instances with $t \geq 3$, they create $2t-4$ additional dummy individuals $\{ a_1^1, \ldots, a_1^{t-2}, a_2^1, \ldots, a_2^{t-2} \}$. They then define the profile $\varphi$ and set $A^+ = \{ a(x,1) : x \in X \} \cup \{ a(C) \} \cup \{ a_1^1, \ldots, a_1^{t-2} \}$.

To turn any instance from the above reduction into a $f^{(s, t)}$-EGCPI instance, we create $s$ additional individuals $\{ d_1, \ldots, d_s \}$. Each of them qualifies everyone (including themselves) but is disqualified by all individuals from the original instance. Since all individuals from the original instance disqualify themselves, their social qualification status is unaffected by the individuals $\{ d_1, \ldots, d_s \}$. We then set $A^- = \{ d_1, \ldots, d_s \} \cup \{ a(x,2) : x \in X \} \cup \{ a(c) : c \in C \} \cup \{ a_2^1, \ldots, a_2^{t-2} \}$.

Observe that $A^- = N \setminus A^+$. Initially the individuals $\{ d_1, \ldots, d_s \}$ are socially qualified because they all qualify themselves and each other (so each of them is qualified by exactly $s$ individuals). However, the attacker can easily turn them all socially disqualified by assigning at least one of them to a different partition than the rest. Furthermore, all individuals from $A^- \setminus \{ d_1, \ldots, d_s \}$ always get eliminated in the first stage of the selection anyway \cite[Theorem 5]{YD18}. Therefore, the requirement that all individuals from $A^-$ must be socially disqualified is fulfilled trivially. The rest of the reduction is equivalent to the constructive case.

From this, it follows that $f^{(s, t)}$-EGCPI is NP-complete for all $s \geq 2$ and $t \geq 2$.
\end{Proof}

When using the consensus-start-respecting rule, EGCAI is NP-complete. This follows from the NP-completeness proof for $f^{\text{CSR}}$-CGCAI by \textcite[Theorem 6]{YD18} where they set $A^+ = T$ (i.e.\@ all individuals must be socially qualified).

\begin{Corollary}
$f^{\text{CSR}}$-\textsc{Exact Group Control by Adding Individuals} is NP-complete.
\end{Corollary}

We now turn to the liberal-start-respecting rule $f^{\text{LSR}}$. From \textcite[Theorem 6]{ERY20}, we already know that this rule is immune to DGCAI. Using similar arguments, we can show that $f^{\text{LSR}}$ is immune to EGCAI when $A^- \neq \emptyset$:

\begin{Theorem} \label{thm:lsr_immune_egcai}
$f^{\text{LSR}}$ is immune to \textsc{Exact Group Control by Adding Individuals}, provided that $A^- \neq \emptyset$.
\end{Theorem}

\begin{Proof}
With the $f^{\text{LSR}}$ rule, it is impossible to turn an individual socially disqualified by adding individuals. To see this, let $a \in A^-$ be an individual who is socially qualified, i.e.\@ $a \in f^{\text{LSR}}(T, \varphi)$. We distinguish between two cases. \\
\begin{tabularx}{\textwidth}{lX}
Case 1: $\varphi(a, a) = 1$ &
In this case, $a \in K^{\text{L}}_0(T, \varphi)$ and therefore $a \in f^{\text{LSR}}(T \cup U, \varphi)$ for all $U \subseteq N \setminus T$ we could add.
\\
Case 2: $\varphi(a, a) = -1$ &
There must exist some sequence of individuals $a_1, a_2, \ldots, a_k \in T$ such that $\varphi(a_1, a_1) = 1$, $\varphi(a_{i}, a_{i+1}) = 1$ for all $i \in \{1, \ldots, k-1\}$, and $\varphi(a_k, a) = 1$. Clearly, this sequence also exists in all supersets of $T$. Thus, $a \in f^{\text{LSR}}(T \cup U, \varphi)$ for all $U \subseteq N \setminus T$ we could add.
\\
\end{tabularx}
\end{Proof}

Note that the above result only applies if $A^- \neq \emptyset$. When $A^- = \emptyset$, $f^{\text{LSR}}$-EGCAI becomes NP-complete. This follows from the NP-completeness proof for $f^{\text{LSR}}$-CGCAI by \textcite[Theorem 6]{YD18} in which they set $A^+ = T$ (i.e.\@ all individuals must be socially qualified).

\begin{Corollary}
$f^{\text{LSR}}$-\textsc{Exact Group Control by Adding Individuals} is NP-complete when $A^- = \emptyset$.
\end{Corollary}

From \textcite[Theorem 7]{YD18}, we know that the liberal-start-respecting rule is immune to both CGCPI and CGCDI. Using similar arguments, we can show that it is immune to EGCPI:

\begin{Theorem} \label{thm:lsr_immune_egcpi}
$f^{\text{LSR}}$ is immune to \textsc{Exact Group Control by Partitioning of Individuals}.
\end{Theorem}

\begin{Proof}
With the $f^{\text{LSR}}$ rule, it is impossible to turn an individual socially qualified by partitioning (or deleting) individuals. To see this, let $a \in A^+$ be an individual who is not socially qualified, i.e.\@ $a \not\in f^{\text{LSR}}(N, \varphi)$. The only way for $a$ to become socially qualified is through a sequence of individuals $a_1, a_2, \ldots, a_k \in N$ such that $\varphi(a_1, a_1) = 1$, $\varphi(a_{i}, a_{i+1}) = 1$ for all $i \in \{1, \ldots, k-1\}$, and $\varphi(a_k, a) = 1$. Clearly, if no such sequence exists in $N$, we cannot hope to create one by partitioning $N$ (or by deleting individuals from $N$).

Note that the above observation only applies if $A^+ \neq \emptyset$. When $A^+ = \emptyset$, we have $A^- = N$, i.e.\@ the task is to make everyone socially disqualified. Provided that not all individuals are socially disqualified initially, there must exist at least one individual $a \in A^-$ who qualifies themselves. Clearly, the individual $a$ would always survive both the first and the second stage of the selection, regardless of how we partition the set. Therefore, the $f^{\text{LSR}}$ rule is immune to EGCPI not only if $A^+ \neq \emptyset$, but even if $A^+ = \emptyset$.
\end{Proof}

\renewcommand{\arraystretch}{1.8}

\subsubsection{Parameterized Complexity}
\label{sec:exact_param}

We now turn to the parameterized complexity of exact group control and bribery problems. Analogous to the constructive case (see \propref{prop:constructive_xp_l}), for all problems with a budget parameter $\ell$, there exist trivial XP-algorithms with respect to $\ell$:

\begin{Proposition} \label{prop:exact_xp_l}
$f$-\textsc{Exact Group Control by Adding Individuals}, $f$-\textsc{Exact Group Bribery} and $f$-\textsc{\$Exact Group Bribery} can be solved in time $\mathcal{O}(n^{\ell+2})$. \\
$f$-\textsc{Exact Group Microbribery} and $f$-\textsc{\$Exact Group Microbribery} can be solved in time $\mathcal{O}(n^{2\ell+2})$.
\end{Proposition}

However, unlike for constructive and destructive cases, note that parameterizing by the target size $|A^+| + |A^-|$ is not useful in the exact case since $|A^+| + |A^-| = |N|$, i.e.\@ the parameter would be equal to the instance size measure.

This leaves the parameters $\ell$, $s$ and $t$ to work with. \textcite[Corollary 4]{BBKL20} show that $f^{(s, t)}$-EGB is W[1]-hard with respect to $\ell + t$ even if $s=1$ and also W[1]-hard with respect to $\ell + s$ even if $t=1$. Naturally, these hardness results also extend to $f^{(s, t)}$-\$EGB. When parameterized by just $s$ and/or $t$, the problem remains para-NP-hard as \textcite[Observation 3]{BBKL20} show that the problem is NP-hard even for $s = 1$ and $t = 2$ (and also for $s = 2$ and $t = 1$).

Assuming $\text{P} \neq \text{NP}$ and $\text{FPT} \neq \text{W[1]}$, the para-NP-hardness and W[1]-hardness results imply that there exists no XP algorithm with respect to $s$ and $t$, and no FPT algorithm for any parameter combination of $\ell$, $s$ and $t$. The only open question is the precise classification of the problems in the W-hierarchy.

\begin{figure}[!htb]
\centering
\begin{minipage}[b]{.45\textwidth}

\begin{tikzpicture}[scale=1] %%%%%%%%%%%%%%%%%%%%%%%%%%%%%%%%%%%%%%%%%%

\spiderweb

% para-NP-hard with respect to s+t
\draw [color=darkgray,line width=1.5pt,opacity=0.5]
    (D1-2) --
    (D2-2) --
    (D3-1) -- cycle;

% W[1]-hard with respect to l+t even when s=1
\draw [color=orange,line width=1.5pt,opacity=0.5]
    (D1-4) --
    (D2-2) --
    (D3-2) -- cycle;

% W[1]-hard with respect to l+s even when t=1
\draw [densely dotted,color=orange,line width=1.5pt,opacity=0.5]
    (D1-2) --
    (D2-4) --
    (D3-2) -- cycle;

% labels for spiderweb
\spiderweblabels

% labels for each dimension axis
\path (1*\A:\L) node (L1) {$s$};
\path (2*\A:\L) node (L2) {$t$};
\path (3*\A:\L) node (L3) {$\ell$};

% color legend
\draw [color=darkgray,line width=1.5pt,opacity=0.5] (2,3) -- (2.4,3)
    node[right,black,opacity=1] {~para-NP-hard};
\draw [color=orange,line width=1.5pt,opacity=0.5] (2,2.5) -- (2.4,2.5);
\draw [densely dotted,color=orange,line width=1.5pt,opacity=0.5] (2,2.3) -- (2.4,2.3)
    node[yshift=0.25em,right,black,opacity=1] {~W[1]-hard};

\end{tikzpicture} %%%%%%%%%%%%%%%%%%%%%%%%%%%%%%%%%%%%%%%%%%

\end{minipage}\hfill
\begin{minipage}[b]{.45\textwidth}

\begin{tikzpicture}[scale=1] %%%%%%%%%%%%%%%%%%%%%%%%%%%%%%%%%%%%%%%%%%

\spiderweb

% XP with respect to l
\draw [color=blue,line width=1.5pt,opacity=0.6]
    (D1-1) --
    (D2-1) --
    (D3-2) -- cycle;

% labels for spiderweb
\spiderweblabels

% labels for each dimension axis
\path (1*\A:\L) node (L1) {$s$};
\path (2*\A:\L) node (L2) {$t$};
\path (3*\A:\L) node (L3) {$\ell$};

% color legend
\draw [color=blue,line width=1.5pt,opacity=0.6] (2,3) -- (2.4,3)
    node[right,black,opacity=1] {~XP};

\end{tikzpicture} %%%%%%%%%%%%%%%%%%%%%%%%%%%%%%%%%%%%%%%%%%

\end{minipage}

\caption{
Parameterized complexity of $f^{(s, t)}$-EGB and -\$EGB for different parameter combinations. The figure on the left shows hardness results and the figure on the right shows results for XP (slice-wise polynomial) algorithms.
}
\label{fig:spiderweb_exact}
\end{figure}
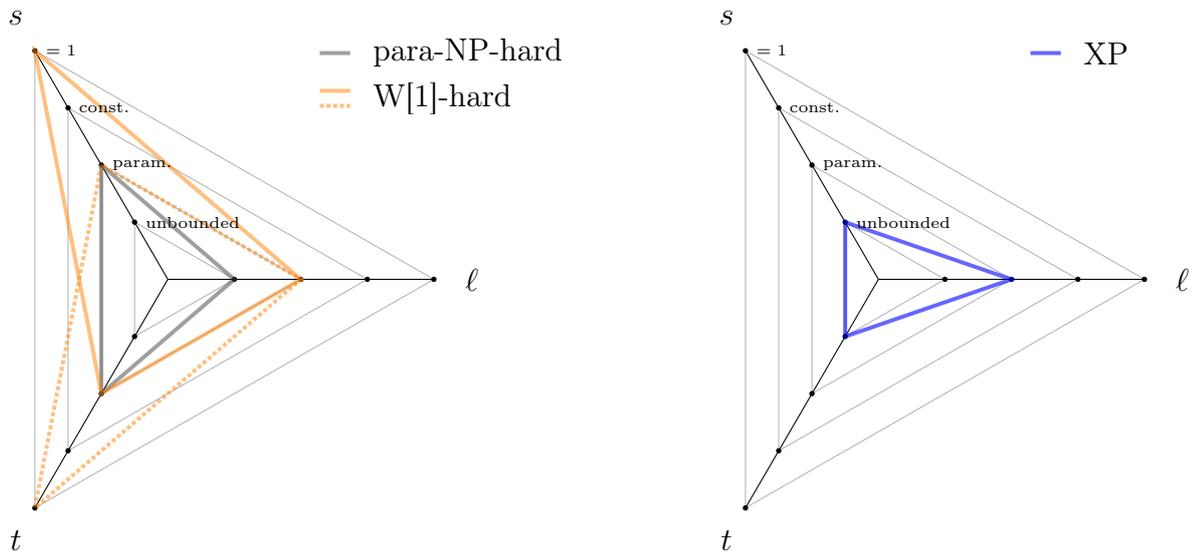

\figref{fig:spiderweb_exact} provides an overview of the parameterized complexity of EGB and \$EGB for the consent rules with different parameter combinations.

In summary, $f^{(s, t)}$-EGB and -\$EGB are

\begin{itemize}[nolistsep]
\item
para-NP-hard with respect to $s+t$
(\tikz\draw [color=darkgray,line width=1.5pt,opacity=0.5] (0,0) -- (0.4,0);, left) \cite[Observation 3]{BBKL20}.

\item
W[1]-hard with respect to $\ell+t$ even when $s=1$
(\tikz\draw [color=orange,line width=1.5pt,opacity=0.5] (0,0) -- (0.4,0);, left) \cite[Corollary 4]{BBKL20}.

\item
W[1]-hard with respect to $\ell+s$ even when $t=1$
(\tikz\draw [densely dotted,color=orange,line width=1.5pt,opacity=0.5] (0,0) -- (0.4,0);, left) \cite[Corollary 4]{BBKL20}.

\item
XP with respect to $\ell$
(\tikz\draw [color=blue,line width=1.5pt,opacity=0.6] (0,0) -- (0.4,0);, right).
See \propref{prop:exact_xp_l}.
\end{itemize}

\clearpage

\clearpage

\subsection{General group control and bribery (constructive+destructive)}
\label{sec:binary_general}

To conclude the section on binary profiles, we now consider the most general form of group control and bribery problems (sometimes referred to as constructive+destructive).

The group control problems GCAI, GCDI and GCPI have not been studied up to now. However, in \secref{sec:general_observations}, we present some easily obtainable results that cover most of the cases. The bribery and microbribery problems were first studied by \textcite{BBKL20}.

Note that, when $A^- = \emptyset$ [resp.\@ $A^+ = \emptyset$], the problems are simply equivalent to the constructive [resp.\@ destructive] cases. For these problem instances, we refer to Tables \ref{tab:binary_constructive} and \ref{tab:binary_destructive}. Below, we only consider instances where both $A^+$ and $A^-$ are nonempty. \tableref{tab:binary_general} lists the known complexity results for group control and bribery restricted to such instances.

\begin{table}[!htb]
\begin{tabularx}{\textwidth}{
p{0.1\textwidth}
X
X
X
X
X
X
X
X
}
\hline & \multicolumn{6}{l}{Consent rules $f^{(s, t)}$} & $f^{\text{CSR}}$ & $f^{\text{LSR}}$ \\
\cline { 2 - 7 } & $s=1$ & $s=1$ & $s \geq 2$ & $s=2$ & $s=2$ & $s \geq 3$ & & \\
\cline { 2 - 7 } & $t=1$ & $t \geq 2$ & $t=1$ & $t=2$ & $t \geq 3$ & $t \geq 2$ & & \\
\hline

\makecell{GCAI \\ ~} &
I &
I &
I &
\makecell{NP-c \\ (+)} &
\makecell{NP-c \\ (+)} &
\makecell{NP-c \\ (+)} &
I &
I \\

\makecell{GCDI \\ ~} &
I &
I &
I &
P &
\makecell{NP-c \\ (+)} &
\makecell{NP-c \\ (+)} &
? &
I \\

\makecell{GCPI \\ ~} &
I &
I &
I &
NP-c &
NP-c &
NP-c &
? &
I \\

\makecell{GB \\ ~} &
P &
\makecell{NP-c \\ (+)} &
\makecell{NP-c \\ (+)} &
\makecell{NP-c \\ (+)} &
\makecell{NP-c \\ (+)} &
\makecell{NP-c \\ (+)} &
P &
P \\

\makecell{\$GB \\ ~} &
P &
\makecell{NP-c \\ (+)} &
\makecell{NP-c \\ (+)} &
\makecell{NP-c \\ (+)} &
\makecell{NP-c \\ (+)} &
\makecell{NP-c \\ (+)} &
P &
P \\

\makecell{GMB \\ ~} &
P &
P &
P &
P &
P &
P &
NP-c &
NP-c \\

\makecell{\$GMB \\ ~} &
P &
P &
P &
P &
P &
P &
NP-c &
NP-c \\

\hline
\end{tabularx}
\caption{
A summary of the complexity results for group control and bribery on instances with $A^+ \neq \emptyset$ and $A^- \neq \emptyset$.
In the table, ``P'' stands for ``polynomial-time solvable'', ``NP-c'' stands for ``NP-complete'', ``I'' stands for ``immune'', and ``?'' means that the complexity of the problem is open.
The symbol ``+'' below a NP-completeness result indicates that the problem is fixed-parameter tractable (FPT) with respect to $|A^+| + |A^-|$.
}
\label{tab:binary_general}
\end{table}

The results for $f^{(s, t)}$-GB and -\$GB are shown by \textcite[Observation 3]{BBKL20}. In particular, the NP-completeness results follow from the fact that $f^{(s, t)}$-CGB and -DGB are NP-complete for all combinations of $s$ and $t$ except $s = t = 1$. Furthermore, \textcite[Corollary 3]{BBKL20} show that $f^{(s, t)}$-GB and -\$GB are FPT with respect to $|A^+| + |A^-|$.

The polynomial-time solvability of $f^{\text{CSR}} / f^{\text{LSR}}$-GB and -\$GB is shown by \textcite[Theorem 1]{BBKL20} via a reduction to \textsc{Minimum Weighted Separator}.

For $f^{(s, t)}$-GMB and -\$GMB, there also exist polynomial-time algorithms as shown by \textcite[Observation 2]{BBKL20}. The algorithms are based on a simple case analysis.

For $f^{\text{CSR}} / f^{\text{LSR}}$-GMB and -\$GMB, \textcite[Theorem 3]{BBKL20} show a reduction from \textsc{Set Cover} to $f^{\text{CSR}} / f^{\text{LSR}}$-CGMB. In the reduction, they create individuals representing the sets and the ground set, as well as one individual $a^\ast$ who is qualified by everyone (including themself). It is easy to turn the resulting CGMB instances into GMB instances with $A^- \neq \emptyset$ by adding one further individual $d$ who qualifies only $a^\ast$, setting $A^- = \{d\}$, letting $a^\ast$ qualify $d$ with $\rho(a^\ast, d) = 1$, and increasing the budget $\ell$ by $1$. Since $a^\ast$ is initially socially qualified, the attacker is always forced to bribe $a^\ast$ to disqualify $d$. This decreases the budget by $1$, so the remaining work is again equivalent to the constructive case. This reduction implies that $f^{\text{CSR}} / f^{\text{LSR}}$-GMB and -\$GMB are NP-complete and W[2]-hard with respect to $\ell$.

In the following section, we present some easily obtainable results for the previously unstudied GCAI, GCDI and GCPI problems.

\subsubsection{Complexity of group control problems}
\label{sec:general_observations}

We begin with the consent rules $f^{(s, t)}$. As we only consider instances where both $A^+$ and $A^-$ are nonempty, the consent rule is immune to GCAI, GCDI and GCPI when either of $s$ or $t$ is set to $1$:

\begin{Corollary} \label{cor:fst_immune_gcai_gcdi_gcpi}
$f^{(s, t)}$ is immune to \textsc{Group Control by Adding Individuals}, \textsc{Group Control by Deleting Individuals} and \textsc{Group Control by Partitioning of Individuals} when $s=1$ or $t=1$, provided that $A^+ \neq \emptyset$ and $A^- \neq \emptyset$.
\end{Corollary}

\begin{Proof}
This follows from the same four facts we already showed in the proofs of \thmref{thm:fst_immune_egcai} and \thmref{thm:fst_immune_egcpi}.
\end{Proof}

When $s \geq 2$ and $t \geq 2$, $f^{(s, t)}$-GCAI is NP-complete:

\begin{Corollary} \label{cor:fst_gcai_npc}
$f^{(s, t)}$-\textsc{Group Control by Adding Individuals} is NP-complete for all $s \geq 2$ and $t \geq 2$.
\end{Corollary}

\begin{Proof}
For all $s \geq 2$ and $t \geq 1$, \textcite[Theorem 4]{YD18} show reductions from \textsc{Restricted Exact Cover by 3-sets} to $f^{(s, t)}$-CGCAI. All resulting CGCAI instances where $t \geq 2$ can easily be turned into GCAI instances with $A^- \neq \emptyset$:

We add $t-1$ individuals $\{ d_1, \ldots, d_{t-1} \}$ to $T$, and one individual $d_t$ to $N \setminus T$. All of these new individuals disqualify everyone, including themselves. All individuals from the original instance qualify $d_1$. We then let $A^- = \{d_1\}$ and increase the budget $\ell$ by $1$.

Note that $d_1$ is initially disqualified by $t-1$ individuals in $T$ (including $d_1$ themself) and is therefore socially qualified. To turn $d_1$ socially disqualified, the attacker is forced to include $d_t$ in the solution $U$ since $d_t$ is the only individual in $N \setminus T$ who disqualifies $d_1$. This decreases the budget by $1$, so the remaining work is again equivalent to the constructive case. From this, it follows that $f^{(s, t)}$-GCAI is NP-complete for all $s \geq 2$ and $t \geq 2$.
\end{Proof}

In contrast to the above, $f^{(s, t)}$-GCDI is still polynomial-time solvable when $s=t=2$:

\begin{Observation}
When $s=2$ and $t=2$, $f^{(s, t)}$-\textsc{Group Control by Deleting Individuals} can be solved in time $\mathcal{O}(n^2)$ where $n = |N|$ denotes the number of individuals.
\end{Observation}

\begin{Proof}
Since $t=2$, any individual in $A^+$ who disqualifies themselves must not be disqualified by anyone else. Let $A^+_{-1} = \{ a \in A^+ : \varphi(a, a) = -1 \}$. We iterate over each $a \in A^+_{-1}$ and delete all individuals who disqualify $a$. We reduce the budget $\ell$ accordingly. This takes time $\mathcal{O}(n^2)$.

Likewise, since $s=2$, any individual in $A^-$ who qualifies themselves must not be qualified by anyone else. Let $A^-_1 = \{ a \in A^- : \varphi(a, a) = 1 \}$. We iterate over each $a \in A^-_1$ and delete all individuals who qualify $a$. We reduce the budget $\ell$ accordingly. This also takes time $\mathcal{O}(n^2)$.

If at any point in the process, the budget is exhausted or we would need to delete an individual from $A^+$ or $A^-$ (which is not allowed), we immediately conclude that the given instance is a NO-instance. Otherwise, we compute the final set of socially qualified individuals and check if all individuals from $A^+$ [resp.\@ $A^-$] are socially qualified [resp.\@ socially disqualified]. If this is the case, we output YES. Else, we output NO as there must exist some individual in $A^+$ [resp.\@ $A^-$] who qualifies [resp.\@ disqualifies] themselves but is qualified [resp.\@ disqualified] by less than $s$ [resp.\@ $t$] individuals; and we cannot hope to change that by deleting more individuals.
\end{Proof}

For all remaining combinations of $s$ and $t$, $f^{(s, t)}$-GCDI is NP-complete:

\begin{Corollary} \label{cor:fst_gcdi_npc}
When $s \geq 2$ and $t \geq 3$, and also when $s \geq 3$ and $t \geq 2$, $f^{(s, t)}$-\textsc{Group Control by Deleting Individuals} is NP-complete.
\end{Corollary}

\begin{Proof}
For all $s \geq 1$ and $t \geq 3$, \textcite[Theorem 4]{YD18} show reductions from \textsc{Restricted Exact Cover by 3-sets} to $f^{(s, t)}$-CGCDI. All resulting CGCDI instances where $s \geq 2$ can easily be turned into GCDI instances with $A^- \neq \emptyset$:

We add $s$ individuals $\{ d_1, \ldots, d_s \}$ and let each of them qualify everyone, including themselves. All individuals from the original instance disqualify $d_1$. We then let $A^- = \{d_1\}$ and increase the budget $\ell$ by $1$.

Note that $d_1$ is initially qualified by $s$ individuals (including $d_1$ themself) and is therefore socially qualified. To turn $d_1$ socially disqualified, the attacker is forced to delete one individual from $\{ d_2, \ldots, d_s \}$. This decreases the budget by $1$, so the remaining work is again equivalent to the constructive case. From this, it follows that $f^{(s, t)}$-GCDI is NP-complete for all $s \geq 2$ and $t \geq 3$.

Using the same reduction in conjunction with \lemmaref{lemma:one}, we can obtain a similar result for all $s \geq 3$ and $t \geq 2$. This concludes the proof.
\end{Proof}

Regarding the $f^{(s, t)}$-GCPI problem, we can use the same reduction as in \thmref{thm:fst_egcpi_npc} to show that it is NP-complete for all $s \geq 2$ and $t \geq 2$:

\begin{Corollary} \label{cor:fst_gcpi_npc}
$f^{(s, t)}$-\textsc{Group Control by Partitioning of Individuals} is NP-complete for all $s \geq 2$ and $t \geq 2$.
\end{Corollary}

\begin{Proof}
This directly follows from \thmref{thm:fst_egcpi_npc} where we show that $f^{(s, t)}$-EGCPI is NP-complete for all $s \geq 2$ and $t \geq 2$. Note that the instances constructed in the proof fulfill the requirement that $A^+ \neq \emptyset$ and $A^- \neq \emptyset$.
\end{Proof}

We now turn to the consensus-start-respecting rule. We have already seen that $f^{\text{CSR}}$-CGCAI and $f^{\text{CSR}}$-DGCAI are NP-complete (see Tables \ref{tab:binary_constructive} and \ref{tab:binary_destructive}). However, when restricted to instances where both $A^+ \neq \emptyset$ and $A^- \neq \emptyset$, the $f^{\text{CSR}}$ rule becomes immune to GCAI. This is a consequence of the fact that we only consider instances where at least one individual from $A^+$ is initially socially disqualified, and at least one individual from $A^-$ is initially socially qualified. Intuitively, it is impossible to turn an individual from $A^-$ socially disqualified and at the same time make all individuals in $A^+$ socially qualified.

\begin{Observation}
$f^{\text{CSR}}$ is immune to \textsc{Group Control by Adding Individuals}, provided that $A^+ \neq \emptyset$ and $A^- \neq \emptyset$.
\end{Observation}

\begin{Proof}
Assume we are given an instance of $f^{\text{CSR}}$-GCAI where $A^+ \neq \emptyset$ and $A^- \neq \emptyset$. Let $G_{N, \varphi}$ be the qualification graph of the given instance. We fix any individual $a^- \in A^-$ who is initially socially qualified, i.e.\@ $a^- \in f^{\text{CSR}}(T, \varphi)$. We also fix any individual $a^+ \in A^+$.

Because $a^-$ is initially socially qualified, there must exist some individual $s \in T$ who is qualified by everyone in $T$, and there must be a path from $s$ to $a^-$ in $G_{N, \varphi}$ that only visits individuals from $T$ (or $s = a^-$ is also possible). The fact that $s$ is qualified by everyone in $T$ implies that $a^+$ also qualifies $s$. Thus, regardless of which individuals are added to $T$, there will always exist a path from $a^+$ via $s$ to $a^-$. Therefore, it is impossible to simultaneously make $a^+$ socially qualified and  $a^-$ socially disqualified.
\end{Proof}

Finally, we turn to the liberal-start-respecting rule. From \thmref{thm:lsr_immune_egcai}, it follows that $f^{\text{LSR}}$ is immune to GCAI:

\begin{Corollary} \label{cor:lsr_immune_gcai}
$f^{\text{LSR}}$ is immune to \textsc{Group Control by Adding Individuals}, provided that $A^- \neq \emptyset$.
\end{Corollary}

\begin{Proof}
As shown in the proof of \thmref{thm:lsr_immune_egcai}, with the $f^{\text{LSR}}$ rule, it is impossible to turn an individual socially disqualified by adding individuals.
\end{Proof}

Similarly, from \thmref{thm:lsr_immune_egcpi}, it follows that $f^{\text{LSR}}$ is immune to GCDI and GCPI:

\begin{Corollary} \label{cor:lsr_immune_gcdi_gcpi}
$f^{\text{LSR}}$ is immune to \textsc{Group Control by Deleting Individuals} and \textsc{Group Control by Partitioning of Individuals}, provided that $A^+ \neq \emptyset$.
\end{Corollary}

\begin{Proof}
As shown in the proof of \thmref{thm:lsr_immune_egcpi}, with the $f^{\text{LSR}}$ rule, it is impossible to turn an individual socially qualified by partitioning or deleting individuals.
\end{Proof}

\subsubsection{Parameterized Complexity}

We now turn to the parameterized complexity of group control and bribery problems. First, we show that $f^{(s, t)}$-GCAI and -GCDI are fixed-parameter tractable with respect to the target size $|A^+| + |A^-|$. To do this, we give ILP formulations for these problems very similar to the ones constructed by \textcite[Theorem 9]{YD18}. The number of variables in the ILP formulations will be bounded by a function of $|A^+| + |A^-|$. This allows us to apply the algorithm by \textcite{L83}.

\begin{Theorem} \label{thm:general_control_fpt_aa}
$f^{(s, t)}$-\textsc{Group Control by Adding Individuals} and $f^{(s, t)}$-\textsc{Group Control by Deleting Individuals} are fixed-parameter tractable with respect to $|A^+| + |A^-|$.
\end{Theorem}

\begin{Proof}
We first consider the GCAI problem.

Let $(N, \varphi, A^+, A^-, T, \ell)$ be an instance of $f^{(s, t)}$-GCAI with $|N| = n$, $|A^+| = q$ and $|A^-| = d$. Let $(a_1, a_2, \ldots, a_n)$ be an arbitrary but fixed order of $N$.
Now we fix a permutation $\lambda$ of the set $A := A^+ \cup A^-$ such that $A = \{ a_{\lambda(1)}, \ldots, a_{\lambda(q)}, a_{\lambda(q+1)}, \ldots, a_{\lambda(q+d)} \}$ where $1 \leq \lambda(i) < \lambda(j) \leq n$ for every $1 \leq i < j \leq q$ and for every $q+1 \leq i < j \leq q+d$. Also, we require that $A^+ = \{ a_{\lambda(1)}, \ldots, a_{\lambda(q)} \}$ and $A^- = \{ a_{\lambda(q+1)}, \ldots, a_{\lambda(q+d)} \}$.

For every $a_i \in N$, let $\varphi(a_i, A)$ denote the vector $\big\langle \varphi(a_i, a_{\lambda(1)}), \varphi(a_i, a_{\lambda(2)}), \ldots, \varphi(a_i, a_{\lambda(q+d)}) \big\rangle$. For every $(q+d)$-dimensional $\{ -1, 1\}$-vector $\beta$, let $N_\beta = \{ a_i \in N \setminus T : \varphi(a_i, A) = \beta \}$ be the set of all individuals we could add whose opinions over $A^+ \cup A^-$ exactly match the vector $\beta$. Let $n_\beta = |N_\beta|$ and let $\beta[i]$ denote the $i$-th component of a vector $\beta$.

For the ILP formulation, let $\mathfrak{B}$ denote the set of all $(q+d)$-dimensional $\{ -1, 1\}$-vectors. For every $\beta \in \mathfrak{B}$, we create a variable $x_\beta$ that indicates how many individuals from $N_\beta$ we include in the solution $U$. This makes for a total of $2^{(q+d)}$ variables. We now formulate the following restrictions:

\begin{enumerate}[nolistsep]
\item[(1)]
For every $\beta \in \mathfrak{B}$, we need to ensure that no less than zero and no more than $n_\beta$ individuals can be included in $U$, thus
$$
0 \leq x_\beta \leq n_\beta.
$$
\item[(2)]
Also, we must ensure that $U$ can consist of at most $\ell$ individuals, thus
$$
\sum_{\beta \in \mathfrak{B}} x_\beta \leq \ell.
$$
\item[(3)]
We need all individuals in $A^+$ to be socially qualified in the final profile:
\subitem[3.1]
For every $a_{\lambda(i)} \in A^+$ where $1 \leq i \leq q$ and $\varphi(a_{\lambda(i)}, a_{\lambda(i)}) = 1$
$$
|T^1_\varphi(a_{\lambda(i)})| + \sum_{\substack{\beta \in \mathfrak{B} \\ \beta[i] = 1}} x_\beta \geq s.
$$
\subitem[3.2]
For every $a_{\lambda(i)} \in A^+$ where $1 \leq i \leq q$ and $\varphi(a_{\lambda(i)}, a_{\lambda(i)}) = -1$
$$
|T^{-1}_\varphi(a_{\lambda(i)})| + \sum_{\substack{\beta \in \mathfrak{B} \\ \beta[i] = -1}} x_\beta \leq t - 1.
$$
\item[(4)]
Also, we need all individuals in $A^-$ to be socially disqualified in the final profile:
\subitem[4.1]
For every $a_{\lambda(i)} \in A^-$ where $q+1 \leq i \leq q+d$ and $\varphi(a_{\lambda(i)}, a_{\lambda(i)}) = 1$
$$
|T^1_\varphi(a_{\lambda(i)})| + \sum_{\substack{\beta \in \mathfrak{B} \\ \beta[i] = 1}} x_\beta \leq s - 1.
$$
\subitem[4.2]
For every $a_{\lambda(i)} \in A^-$ where $q+1 \leq i \leq q+d$ and $\varphi(a_{\lambda(i)}, a_{\lambda(i)}) = -1$
$$
|T^{-1}_\varphi(a_{\lambda(i)})| + \sum_{\substack{\beta \in \mathfrak{B} \\ \beta[i] = -1}} x_\beta \geq t.
$$
\end{enumerate}

Recall that $T^1_\varphi(a)$ [resp.\@ $T^{-1}_\varphi(a)$] denotes the set of individuals who qualify [resp.\@ disqualify] the individual $a$. The inequality [3.1] ensures that all $a \in A^+$ who qualify themselves are qualified by at least $s$ individuals, and the inequality [3.2] ensures that all $a \in A^+$ who disqualify themselves are disqualified by at most $t-1$ individuals in the final profile. Likewise, the inequality [4.1] ensures that all $a \in A^-$ who qualify themselves are qualified by at most $s-1$ individuals, and the inequality [4.2] ensures that all $a \in A^-$ who disqualify themselves are disqualified by at least $t$ individuals in the final profile.

Now we consider the GCDI problem.

Let $(N, \varphi, A^+, A^-, \ell)$ be an instance of $f^{(s, t)}$-GCDI with $|N| = n$, $|A^+| = q$ and $|A^-| = d$. We give an ILP formulation similar to the one shown above for GCAI. This time, for every $\beta \in \mathfrak{B}$, we use $\overline{N}_\beta = \{ a_i \in N \setminus A : \varphi(a_i, A) = \beta \}$ to denote the set of all individuals we could delete whose opinions over $A^+ \cup A^-$ exactly match the vector $\beta$. Let $\overline{n}_\beta = |\overline{N}_\beta|$. For every $\beta \in \mathfrak{B}$, we create a variable $y_\beta$ that indicates how many individuals from $\overline{N}_\beta$ we delete. Thus, we again have $2^{(q+d)}$ variables in total. We then formulate the following restrictions:

\begin{enumerate}[nolistsep]
\item[(1)]
For every $\beta \in \mathfrak{B}$, we need to ensure that no less than zero and no more than $\overline{n}_\beta$ individuals can be deleted, thus
$$
0 \leq y_\beta \leq \overline{n}_\beta.
$$
\item[(2)]
Also, we can only delete at most $\ell$ individuals in total, thus
$$
\sum_{\beta \in \mathfrak{B}} y_\beta \leq \ell.
$$
\item[(3)]
We need all individuals in $A^+$ to be socially qualified in the final profile:
\subitem[3.1]
For every $a_{\lambda(i)} \in A^+$ where $1 \leq i \leq q$ and $\varphi(a_{\lambda(i)}, a_{\lambda(i)}) = 1$
$$
|N^1_\varphi(a_{\lambda(i)})| - \sum_{\substack{\beta \in \mathfrak{B} \\ \beta[i] = 1}} y_\beta \geq s.
$$
\subitem[3.2]
For every $a_{\lambda(i)} \in A^+$ where $1 \leq i \leq q$ and $\varphi(a_{\lambda(i)}, a_{\lambda(i)}) = -1$
$$
|N^{-1}_\varphi(a_{\lambda(i)})| - \sum_{\substack{\beta \in \mathfrak{B} \\ \beta[i] = -1}} y_\beta \leq t - 1.
$$
\item[(4)]
Also, we need all individuals in $A^-$ to be socially disqualified in the final profile:
\subitem[4.1]
For every $a_{\lambda(i)} \in A^-$ where $q+1 \leq i \leq q+d$ and $\varphi(a_{\lambda(i)}, a_{\lambda(i)}) = 1$
$$
|N^1_\varphi(a_{\lambda(i)})| - \sum_{\substack{\beta \in \mathfrak{B} \\ \beta[i] = 1}} y_\beta \leq s - 1.
$$
\subitem[4.2]
For every $a_{\lambda(i)} \in A^-$ where $q+1 \leq i \leq q+d$ and $\varphi(a_{\lambda(i)}, a_{\lambda(i)}) = -1$
$$
|N^{-1}_\varphi(a_{\lambda(i)})| - \sum_{\substack{\beta \in \mathfrak{B} \\ \beta[i] = -1}} y_\beta \geq t.
$$
\end{enumerate}

For both ILP formulations, the number of variables is bounded by a function of $|A^+| + |A^-|$. Hence, we can conclude that $f^{(s, t)}$-GCAI and $f^{(s, t)}$-GCDI are FPT with respect to $|A^+| + |A^-|$ \cite{L83,K87,FT87}.
\end{Proof}

Analogous to the constructive case (see \propref{prop:constructive_xp_l}), for all problems with a budget parameter $\ell$, there exist trivial XP-algorithms with respect to $\ell$:

\begin{Proposition} \label{prop:general_xp_l}
The problems $f$-\textsc{Group Control by Adding Individuals}, $f$-\textsc{Group Control by Deleting Individuals}, $f$-\textsc{Group Bribery} and $f$-\textsc{\$Group Bribery} can be solved in time $\mathcal{O}(n^{\ell+2})$. \\
$f$-\textsc{Group Microbribery} and $f$-\textsc{\$Group Microbribery} can be solved in time $\mathcal{O}(n^{2\ell+2})$.
\end{Proposition}

Below, we present the proofs for $f^{\text{LSR}}$-\textsc{\$Group Bribery} and $f^{\text{LSR}}$-\textsc{\$Group Microbribery}. The proofs for the other problems work similarly.

\begin{Proof} \textbf{($f^{\text{LSR}}$-\textsc{\$GB})} ~
Given an $f^{\text{LSR}}$-\$GB instance $(N, \varphi, A^+, A^-, \rho, \ell)$ with $n = |N|$, we guess the subset of individuals we must bribe to make everyone $A^+$ socially qualified and everyone in $A^-$ socially disqualified. To do that, we iterate over all subsets $U \subseteq N$ of size $|U| \leq \ell$.

For each subset $U \subseteq N$, we obtain an altered profile $\varphi^\prime$ by bribing the individuals in $U$ to qualify themselves and everyone in $A^+$ and to disqualify all other individuals. We then check if $\rho(U) \leq \ell$, $A^+ \subseteq f^{\text{LSR}}(N, \varphi^\prime)$, and $A^- \cap f^{\text{LSR}}(N, \varphi^\prime) = \emptyset$. If we find a subset $U$ that satisfies these conditions, the instance is a YES-instance. Otherwise, we know that bribing any group of up to $\ell$ individuals is not sufficient to achieve the strategic objective. Since $\rho$ assigns a positive integer price to each individual, it follows that bribing any group of individuals with cost at most $\ell$ is not sufficient to achieve the strategic objective. Therefore, the instance must be a NO-instance.

The checks for any subset $U \subseteq N$ can be done in time $\mathcal{O}(n^2)$. As there are $\binom{n}{\leq \ell} \in \mathcal{O}(n^\ell)$ subsets to consider, we get a total running time of $\mathcal{O}(n^{\ell+2})$.
\end{Proof}

\begin{Proof} \textbf{($f^{\text{LSR}}$-\textsc{\$GMB})} ~
Given an $f^{\text{LSR}}$-\$GMB instance $(N, \varphi, A^+, A^-, \rho, \ell)$ with $n = |N|$, we guess the subset of pairs of individuals whose valuations we need to change. To do that, we iterate over all subsets $M \subseteq N \times N$ of size $|M| \leq \ell$.

For each subset $M \subseteq N \times N$, we obtain an altered profile $\varphi^\prime$ by flipping the valuations for all pairs $(a, b) \in M$, i.e.\@ if $\varphi(a, b) = 1$ then $\varphi^\prime(a, b) = -1$, and if $\varphi(a, b) = -1$ then $\varphi^\prime(a, b) = 1$. We then check if $\rho(M) \leq \ell$, $A^+ \subseteq f^{\text{LSR}}(N, \varphi^\prime)$, and $A^- \cap f^{\text{LSR}}(N, \varphi^\prime) = \emptyset$. If we find a subset $M$ that satisfies these conditions, the instance is a YES-instance. Otherwise, we know that flipping the valuations for any set of up to $\ell$ pairs is not sufficient to achieve the strategic objective. Since $\rho$ assigns a positive integer price to each pair, it follows that changing the valuations for any set of pairs with cost at most $\ell$ is not sufficient to achieve the strategic objective. Therefore, the instance must be a NO-instance.

The checks for any subset $M \subseteq N \times N$ can be done in time $\mathcal{O}(n^2)$. As there are $\binom{n^2}{\leq \ell} \in \mathcal{O}(n^{2\ell})$ subsets to consider, we get a total running time of $\mathcal{O}(n^{2\ell+2})$.
\end{Proof}

Finally, we look at combinations of the parameters $\ell$, $s$ and $t$. \textcite[Corollary 3]{BBKL20} show that $f^{(s, t)}$-GB is W[1]-hard with respect to $\ell + t$ even if $s=1$ and also W[1]-hard with respect to $\ell + s$ even if $t=1$. Naturally, these hardness results also extend to $f^{(s, t)}$-\$GB. When parameterized by just $s$ and/or $t$, the problem remains para-NP-hard as \textcite[Observation 3]{BBKL20} show that the problem is NP-hard even for $s = 1$ and $t = 2$ (and also for $s = 2$ and $t = 1$). Assuming $\text{P} \neq \text{NP}$ and $\text{FPT} \neq \text{W[1]}$, this effectively rules out any further XP or FPT algorithms with these parameters (as we discussed in \secref{sec:exact_param}).

\figref{fig:spiderweb_exact} provides an overview of the parameterized complexity of GB and \$GB for the consent rules with different parameter combinations.

In summary, $f^{(s, t)}$-GB and -\$GB are

\begin{itemize}[nolistsep]
\item
para-NP-hard with respect to $s+t$
(\tikz\draw [color=darkgray,line width=1.5pt,opacity=0.5] (0,0) -- (0.4,0);, left) \cite[Observation 3]{BBKL20}.

\item
W[1]-hard with respect to $\ell+t$ even when $s=1$
(\tikz\draw [color=orange,line width=1.5pt,opacity=0.5] (0,0) -- (0.4,0);, left) \cite[Corollary 3]{BBKL20}.

\item
W[1]-hard with respect to $\ell+s$ even when $t=1$
(\tikz\draw [densely dotted,color=orange,line width=1.5pt,opacity=0.5] (0,0) -- (0.4,0);, left) \cite[Corollary 3]{BBKL20}.

\item
XP with respect to $\ell$
(\tikz\draw [color=blue,line width=1.5pt,opacity=0.6] (0,0) -- (0.4,0);, right).
See \propref{prop:general_xp_l}.
\end{itemize}

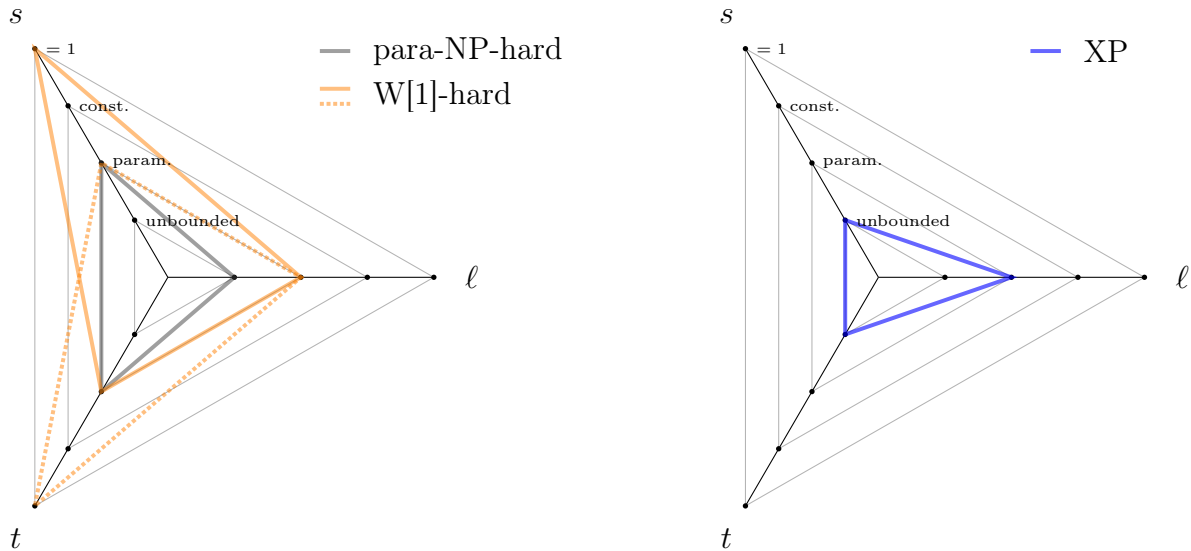
\begin{figure}[!htb]
\centering
\begin{minipage}[b]{.45\textwidth}

\begin{tikzpicture}[scale=1] %%%%%%%%%%%%%%%%%%%%%%%%%%%%%%%%%%%%%%%%%%

\spiderweb

% para-NP-hard with respect to s+t
\draw [color=darkgray,line width=1.5pt,opacity=0.5]
    (D1-2) --
    (D2-2) --
    (D3-1) -- cycle;

% W[1]-hard with respect to l+t even when s=1
\draw [color=orange,line width=1.5pt,opacity=0.5]
    (D1-4) --
    (D2-2) --
    (D3-2) -- cycle;

% W[1]-hard with respect to l+s even when t=1
\draw [densely dotted,color=orange,line width=1.5pt,opacity=0.5]
    (D1-2) --
    (D2-4) --
    (D3-2) -- cycle;

% labels for spiderweb
\spiderweblabels

% labels for each dimension axis
\path (1*\A:\L) node (L1) {$s$};
\path (2*\A:\L) node (L2) {$t$};
\path (3*\A:\L) node (L3) {$\ell$};

% color legend
\draw [color=darkgray,line width=1.5pt,opacity=0.5] (2,3) -- (2.4,3)
    node[right,black,opacity=1] {~para-NP-hard};
\draw [color=orange,line width=1.5pt,opacity=0.5] (2,2.5) -- (2.4,2.5);
\draw [densely dotted,color=orange,line width=1.5pt,opacity=0.5] (2,2.3) -- (2.4,2.3)
    node[yshift=0.25em,right,black,opacity=1] {~W[1]-hard};

\end{tikzpicture} %%%%%%%%%%%%%%%%%%%%%%%%%%%%%%%%%%%%%%%%%%

\end{minipage}\hfill
\begin{minipage}[b]{.45\textwidth}

\begin{tikzpicture}[scale=1] %%%%%%%%%%%%%%%%%%%%%%%%%%%%%%%%%%%%%%%%%%

\spiderweb

% XP with respect to l
\draw [color=blue,line width=1.5pt,opacity=0.6]
    (D1-1) --
    (D2-1) --
    (D3-2) -- cycle;

% labels for spiderweb
\spiderweblabels

% labels for each dimension axis
\path (1*\A:\L) node (L1) {$s$};
\path (2*\A:\L) node (L2) {$t$};
\path (3*\A:\L) node (L3) {$\ell$};

% color legend
\draw [color=blue,line width=1.5pt,opacity=0.6] (2,3) -- (2.4,3)
    node[right,black,opacity=1] {~XP};

\end{tikzpicture} %%%%%%%%%%%%%%%%%%%%%%%%%%%%%%%%%%%%%%%%%%

\end{minipage}

\caption{
Parameterized complexity of $f^{(s, t)}$-GB and -\$GB for different parameter combinations. The figure on the left shows hardness results and the figure on the right shows results for XP (slice-wise polynomial) algorithms.
}
\label{fig:spiderweb_general}
\end{figure}

\clearpage

\section{Restriction to \texorpdfstring{$r$-profiles}{}}
\label{sec:r_profiles}

In this section, we study the complexity of bribery and control problems when they are restricted to $r$-profiles:

An $r$-profile over $N$ is a binary profile over $N$ where each individual $a \in N$ qualifies exactly $r$ individuals (for some positive integer $r$).

In principle, all problems listed in \secref{sec:problem_definitions} can be restricted to $r$-profiles and studied for each of the social rules defined in \secref{sec:social_rules}. Restricting a problem to $r$-profiles can change its complexity. However, note that if a social rule is immune to an attack in general, then it is also immune when restricted to $r$-profiles.

For group control by adding individuals, it is unclear what should happen when some individual $a \in N$ qualifies some other individual $b \in N \setminus T$ but the individual $b$ is not included in $U$ (where $U \subseteq N \setminus T$ is the set of individuals we add to $T$). In this situation, the requirement that each individual must qualify exactly $r$ individuals would not be met after adding $U$ to $T$. To resolve this, alternative problem definitions would be required. However, in \secref{sec:cgcai_r_profiles}, we work around the issue by constructing profiles where all individuals only qualify themselves or individuals in $T$. This way, our results hold regardless of the chosen definition.

When considering group control by deleting individuals, if some individual $a \in N$ is qualified by some other individual $b \in N$ but the individual $a$ is later deleted, $b$ would no longer qualify exactly $r$ individuals. Therefore, when restricted to $r$-profiles, the problems would need to be redefined. For example, we could force the attacker to also delete all individuals qualifying $a$ each time they delete an individual $a \in N$. However, note that this definition could change the complexity of the problems as it restricts the solution space.

For group control by partitioning, the restriction to $r$-profiles raises the question who an individual would qualify if we put them in a different partition than some individuals they originally qualified. Assuming we would uphold the requirement that each individual must qualify exactly $r$ individuals within their assigned partition, it is not clear how to handle that. Therefore, when group control by partitioning of individuals is restricted to $r$-profiles, it turns into a different problem with a potentially different complexity. One possible solution would be to let each individual provide a ranked list of choices over all individuals in the set. Then, when the individuals are assigned to partitions, each individual would qualify exactly those $r$ individuals within their partition who have the highest ranking in the respective list. Obviously, this would require both partitions to consist of at least $r$ individuals.

In group bribery and microbribery problems, note that the attacker must maintain the $r$-profile even after the bribery. Because of this restriction, group bribery and microbribery problems can have a different complexity when restricted to $r$-profiles.

Below, we provide an overview of the known complexity results for CGMB and CGCAI when restricted to $r$-profiles. It is possible that complexity results for other problems listed in \secref{sec:problem_definitions} (when restricted to $r$-profiles) could be obtained in a similar manner. We leave these problems open for future research.

\subsection{Constructive group microbribery}
\label{sec:cgmb_r_profiles}

\textcite{EY20} study the \textsc{Constructive Group Microbribery} problem restricted to $r$-profiles. \tableref{tab:cgmb_r_profiles} provides an overview of their results.

\begin{table}[!htb]
\begin{tabularx}{\textwidth}{
p{0.12\textwidth}
X
X
X
X
X
X
}
\hline CGMB & \multicolumn{4}{l}{Consent rules $f^{(s, t)}$} & $f^{\text{CSR}}$ & $f^{\text{LSR}}$ \\
\cline { 2 - 5 } & $s=1$ & & $s \geq 2$ & & & \\
\cline { 2 - 5 } & $t=1$ & $t \geq 2$ & $t=1$ & $t \geq 2$ & & \\
\hline

$r=1$ &
$\mathcal{O}(n)$ &
$\mathcal{O}(n^2)$ &
$\mathcal{O}(n^2)$ &
? &
$\mathcal{O}(n)$ &
$\mathcal{O}(n)$ \\

$r=2$ &
? &
? &
? &
? &
? &
? \\

$r=3$ &
? &
? &
? &
? &
? &
NP-c \\

$r \geq 4$ &
? &
? &
? &
? &
NP-c &
NP-c \\

\hline
\end{tabularx}
\caption{
A summary of the complexity results for CGMB when restricted to $r$-profiles.
In the table, $n$ denotes the number of individuals, ``NP-c'' stands for ``NP-complete'', and ``?'' means that the complexity of the problem is open.
}
\label{tab:cgmb_r_profiles}
\end{table}

The polynomial-time results for the consent rules are by \textcite[Theorem 7]{EY20}. They only apply if either of $s$ or $t$ is set to $1$, and they only cover the case $r=1$.

For the CSR rule, the problem is linear-time solvable when $r=1$ \cite[Theorem 10]{EY20} and NP-complete for all $r \geq 4$ \cite[Theorem 11]{EY20}. The NP-completeness is shown by a reduction from \textsc{Exact Cover by 3-sets}.

For the LSR rule, the problem is also linear-time solvable when $r=1$ \cite[Theorem 8]{EY20} and NP-complete for all $r \geq 3$ \cite[Theorem 9]{EY20}. The NP-completeness is again shown by a reduction from \textsc{Exact Cover by 3-sets}.

When $r=1$, every individual either qualifies themselves or some other individual. Therefore, the $f^{\text{LSR}}$ rule and the liberal rule $f^{(1, 1)}$ are equivalent in this case (all individuals who qualify themselves are socially qualified, and all other individuals are socially disqualified). Thus, the linear-time result for $f^{\text{LSR}}$-CGMB also applies to $f^{(1, 1)}$-CGMB when $r=1$.

\subsection{Constructive group control by adding individuals}
\label{sec:cgcai_r_profiles}

Using similar approaches as \textcite{EY20}, we can obtain many complexity results for CGCAI restricted to $r$-profiles. \tableref{tab:cgcai_r_profiles} provides an overview of our results.

\begin{table}[!htb]
\begin{tabularx}{\textwidth}{
p{0.12\textwidth}
X
X
X
X
X
X
}
\hline CGCAI & \multicolumn{4}{l}{Consent rules $f^{(s, t)}$} & $f^{\text{CSR}}$ & $f^{\text{LSR}}$ \\
\cline { 2 - 5 } & $s=1$ & & $s \geq 2$ & & & \\
\cline { 2 - 5 } & $t=1$ & $t \geq 2$ & $t=1$ & $t \geq 2$ & & \\
\hline

$r=1$ &
I &
I &
$\mathcal{O}(n^2)$ &
$\mathcal{O}(n^2)$ &
I &
I \\

$r=2$ &
I &
I &
? &
? &
? &
? \\

$r=3$ &
I &
I &
NP-c &
NP-c &
? &
? \\

$r \geq 4$ &
I &
I &
NP-c &
NP-c &
? &
NP-c \\

\hline
\end{tabularx}
\caption{
A summary of the complexity results for CGCAI when restricted to $r$-profiles.
In the table, $n$ denotes the number of individuals, ``NP-c'' stands for ``NP-complete'', ``I'' stands for ``immune'', and ``?'' means that the complexity of the problem is open.
}
\label{tab:cgcai_r_profiles}
\end{table}

The immunity of the consent rules when $s=1$ follows directly from the immunity for general profiles (see \tableref{tab:binary_constructive}). Since the liberal rule $f^{(1, 1)}$ and the $f^{\text{LSR}}$ rule are equivalent when $r=1$, the immunity also extends to the $f^{\text{LSR}}$ rule in this case.

Moving to the $f^{\text{CSR}}$ rule, when $r=1$ the set of socially qualified individuals is either empty or consists of exactly one individual. Hence, if $|A^+| > 1$ it is impossible to make all individuals in $A^+$ socially qualified. But even when $A^+$ consists of just one individual, provided that this individual is not initially qualified by all individuals, it is impossible to turn the individual socially qualified by adding more individuals. Therefore, the $f^{\text{CSR}}$ rule is immune to CGCAI when restricted to $r$-profiles with $r=1$.

Next, we show that $f^{(s, t)}$-CGCAI restricted to $r$-profiles is polynomial-time solvable when $r=1$ for all $s \geq 2$ and $t \geq 1$:

\begin{Theorem} \label{thm:fst_cgcai_r_profiles_p}
$f^{(s, t)}$-\textsc{Constructive Group Control by Adding Individuals} restricted to $r$-profiles can be solved in time $\mathcal{O}(n^2)$ when $r=1$ for all $s \geq 2$ and $t \geq 1$.
\end{Theorem}

\begin{Proof}
Given the $f^{(s, t)}$-CGCAI instance $(N, \varphi, A^+, T, \ell)$, we first compute $A^+_{-1} = \{ a \in A^+ : \varphi(a, a) = -1 \}$, i.e.\@ the set of individuals in $A^+$ who disqualify themselves. For each $a \in A^+_{-1}$, we then check whether $|T^{-1}_\varphi(a)| \geq t$. If we find an individual $a \in A^+_{-1}$ where this is the case, we immediately conclude that the given instance is a NO-instance. The reason for this is that $a$ disqualifies themselves and is disqualified by at least $t$ individuals in $T$. Hence, regardless of which individuals we add to $T$, it is impossible to turn $a$ socially qualified.

Otherwise, we compute $d_a = t - |T^{-1}_\varphi(a)| - 1$ for each individual $a \in A^+_{-1}$. The value $d_a$ denotes the maximum number of additional disqualifications $a$ is allowed to get before they would turn socially disqualified. We let
\renewcommand{\arraystretch}{1}
$$
d = \left\{\begin{array}{ll}
    \operatorname{min}_{a \in A^+_{-1}} (d_a) & \text { if } A^+_{-1} \neq \emptyset \\
    \ell & \text { else. }
    \end{array}\right.
$$
\renewcommand{\arraystretch}{1.8}

Now, we take care of the individuals in $A^+ \setminus A^+_{-1}$, i.e.\@ the set of individuals in $A^+$ who qualify themselves. Each such individual needs at least $s$ qualifications to be socially qualified. Hence, for each $a \in A^+ \setminus A^+_{-1}$, we must add $q_a = \operatorname{max}{(0, s - |T^1_\varphi(a)|)}$ individuals who qualify $a$. Because $r=1$, every individual in $N \setminus T$ qualifies at most one individual in $T$. Hence, each $a \in A^+ \setminus A^+_{-1}$ can be handled separately.

More precisely, for each $a \in A^+ \setminus A^+_{-1}$, we check if there exist at least $q_a$ individuals in $N \setminus T$ who qualify $a$. If we find any $a \in A^+ \setminus A^+_{-1}$ for which this is not the case, we conclude that the instance is a NO-instance. Otherwise, we add the $q_a$ individuals to $T$, update $\ell := \ell - q_a$ and $d := d - q_a$ and proceed with the next $a \in A^+ \setminus A^+_{-1}$.

Once this is done, we perform two more checks: If $\ell < 0$, we output NO (budget exceeded). If $\ell \geq 0$ but $d < 0$, we also output NO (some individual in $A^+_{-1}$ turned socially disqualified because we added too many individuals). If $\ell \geq 0$ and $d \geq 0$, we output YES.

It remains to analyse the running time of the algorithm. The set $A^+_{-1}$ can be computed in time $\mathcal{O}(n)$. All checks regarding the qualification profile $\varphi$ can be performed in time $\mathcal{O}(n^2)$. The values $d_a$, $d$ and $q_a$ can also be computed in time $\mathcal{O}(n^2)$. Processing all individuals in $A^+ \setminus A^+_{-1}$ takes time $\mathcal{O}(n \cdot n)$. Overall, the running time is bounded by $\mathcal{O}(n^2)$.
\end{Proof}

The NP-completeness of $f^{(s, t)}$-CGCAI restricted to $r$-profiles with $r \geq 3$ for all $s \geq 2$ and $t \geq 1$ can be shown by a simple reduction from \textsc{Restricted Exact Cover by 3-sets} (RX3C). The reduction is similar to the one by \textcite{EY20} as well as the one by \textcite{YD18}:

\begin{Theorem} \label{thm:fst_cgcai_r_profiles_npc}
$f^{(s, t)}$-\textsc{Constructive Group Control by Adding Individuals} restricted to $r$-profiles is NP-complete for all $r \geq 3$, $s \geq 2$ and $t \geq 1$.
\end{Theorem}

\begin{Proof}
We first consider the case where $r=3$ and $s=2$. Given a RX3C-instance $(X, \mathcal{F})$ with $|X|=3m$, we construct an instance of CGCAI as follows:

We introduce one individual $a_x$ for each element $x \in X$ and let $N_X = \{a_x : x \in X\}$. We also introduce one individual $a_F$ for each triplet $F \in \mathcal{F}$ and let $N_\mathcal{F} = \{a_F : F \in \mathcal{F}\}$. Finally, we introduce three dummy individuals $N_D = \{d_1, d_2, d_3\}$. We let $N = N_X \cup N_\mathcal{F} \cup N_D$.

For each $x \in X$, we let $a_x$ qualify only themselves and two arbitrary but fixed dummy individuals. For each $F \in \mathcal{F}$, we let $a_F$ qualify only the three individuals $a_x \in N_X$ for which $x \in F$. We let each dummy individual qualify $d_1$, $d_2$ and $d_3$. Finally, we set $T = N_X \cup N_D$, $A^+ = N_X$, $s=2$, and $\ell=m$. Note that each individual $a_x \in N_X$ is initially only qualified by themselves and therefore needs one more qualification to become socially qualified.

Clearly, the qualification profile in the constructed instance is an $r$-profile with $r=3$. We now show that the constructed instance is a YES-instance if and only if there exists an exact 3-set cover for $X$ in $\mathcal{F}$.

($\Rightarrow$) Assume that $\mathcal{F}^\prime \subseteq \mathcal{F}$ is an exact 3-set cover for $X$. Obviously, the size of $\mathcal{F}^\prime$ is $m = \ell$. By adding all individuals $a_F$ where $F \in \mathcal{F}^\prime$ to $T$, each individual $a_x \in N_X$ gains exactly one qualification. Thus, all individuals in $A^+$ are now socially qualified.

($\Leftarrow$) Assume we are given a set $U \subseteq N \setminus T = N_\mathcal{F}$ of size at most $\ell$ such that all individuals in $A^+$ become socially qualified after adding $U$ to $T$. Because $s=2$, each individual in $N_X$ must have gained at least one qualification by adding $U$ to $T$. From this it follows that the size of $U$ is exactly $m$ and $\mathcal{F}^\prime = \{F \in \mathcal{F} : a_F \in U\}$ is an exact 3-set cover for $X$.

It is easy to extend this proof to work for larger values of $r$ by adding more dummy individuals and pointing additional qualifications into $N_D$. For larger values of $s$, we add further dummy individuals who provide each individual in $A^+$ exactly $s-2$ additional qualifications.
\end{Proof}

By another reduction from RX3C, we can also show that $f^{\text{LSR}}$-CGCAI restricted to $r$-profiles with $r \geq 4$ is NP-complete:

\begin{Theorem} \label{thm:lsr_cgcai_r_profiles_npc}
$f^{\text{LSR}}$-\textsc{Constructive Group Control by Adding Individuals} restricted to $r$-profiles is NP-complete for all $r \geq 4$.
\end{Theorem}

\begin{Proof}
We first consider the case where $r=4$. Given a RX3C-instance $(X, \mathcal{F})$ with $|X|=3m$, we construct an instance of CGCAI as follows:

We introduce one individual $a_x$ for each element $x \in X$ and let $N_X = \{a_x : x \in X\}$. We also introduce one individual $a_F$ for each triplet $F \in \mathcal{F}$ and let $N_\mathcal{F} = \{a_F : F \in \mathcal{F}\}$. Finally, we introduce four dummy individuals $N_D = \{d_1, d_2, d_3, d_4\}$. We let $N = N_X \cup N_\mathcal{F} \cup N_D$.

For each $x \in X$, we let $a_x$ qualify only the four dummy individuals. For each $F \in \mathcal{F}$, we let $a_F$ qualify only themselves and the three individuals $a_x \in N_X$ for which $x \in F$. We let each dummy individual qualify $d_1$, $d_2$, $d_3$ and $d_4$. We then set $T = N_X \cup N_D$, $A^+ = N_X$, and $\ell=m$. Note that we initially have $N_X \cap f^{\text{LSR}}(T, \varphi) = \emptyset$, i.e.\@ none of the individuals in $N_X$ are socially qualified. But since each individual in $N_\mathcal{F}$ qualifies themselves, when we add an individual from $N_\mathcal{F}$ to $T$, the three corresponding individuals from $N_X$ turn socially qualified.

Clearly, the qualification profile in the constructed instance is an $r$-profile with $r=4$. We now show that the constructed instance is a YES-instance if and only if there exists an exact 3-set cover for $X$ in $\mathcal{F}$.

($\Rightarrow$) Assume that $\mathcal{F}^\prime \subseteq \mathcal{F}$ is an exact 3-set cover for $X$. Obviously, the size of $\mathcal{F}^\prime$ is $m = \ell$. We let $U = \{ a_F : F \in \mathcal{F}^\prime \}$. Since $\mathcal{F}^\prime$ is an exact 3-set cover for $X$, each individual in $N_X$ is qualified by some individual in $U$. Therefore, by adding the individuals in $U$ to $T$, all individuals in $A^+$ become socially qualified.

($\Leftarrow$) Assume we are given a set $U \subseteq N \setminus T = N_\mathcal{F}$ of size at most $\ell$ such that all individuals in $A^+$ become socially qualified after adding $U$ to $T$. By construction, each individual in $N_X$ must be qualified by some individual in $U$. Because each individual in $U$ qualifies exactly three individuals in $N_X$, it follows that the size of $U$ is exactly $m$. Thus, $\mathcal{F}^\prime = \{F \in \mathcal{F} : a_F \in U\}$ is an exact 3-set cover for $X$.

It is easy to extend this proof to work for larger values of $r$ by adding more dummy individuals and pointing additional qualifications into $N_D$.
\end{Proof}

The complexity of the remaining cases remains open. In particular, this includes most CGCAI instances with $r=2$ or $r=3$. The results by \textcite{EY20} for CGMB (where they also leave most of these cases open) suggest that analyzing the missing cases could be difficult. Furthermore, to obtain results for the $f^{\text{CSR}}$ rule, we would likely need to create instances where some individuals also qualify individuals in $N \setminus T$ (because it is otherwise not possible to utilize the individuals from $N \setminus T$ in any way). To do this, we would first have to answer the question what should happen when an individual from $N \setminus T$ is qualified by someone but is not included in the solution. In other words, an alternative problem definition would be required, e.g.\@ one where individuals provide a ranked list of choices.

\clearpage

\section{Ternary profiles (with indifference)}
\label{sec:ternary}

So far, we have only considered profiles where every individual holds an opinion about every other individual. There are many scenarios where this assumption may not be realistic. For example, an individual could be indifferent about the qualification status of some other individual, or they might not be familiar with certain individuals in the set, or the given set of individuals could be so large that it is not possible for all individuals to provide their full valuations. To deal with scenarios like this, \textcite[Section 6]{ERY20} introduce and study an extension of the canonical group identification model. In this section, we provide an overview of their results and show how to extend them to the general and exact problem cases.

A ternary profile over $N$ is a function $\varphi : N \times N \rightarrow \{ -1, \star, 1 \}$. For $a,b \in N$, we write $\varphi(a,b) = \star$ to denote that $a$ is indifferent whether $b$ is qualified or not.

We use social rules similar to the consent rules to determine the set of socially qualified individuals for a given ternary profile over a set $N$. Each such rule is characterized by three parameters $s, s^\prime, t \in \mathbb{N}$ and denoted as $f^{(s, s^\prime, t)}$. For every subset $T \subseteq N$ and every individual $a \in N$, it holds:
\setlist[itemize,1]{label=\normalfont\bfseries\textendash}
\begin{itemize}[nolistsep,topsep=-4pt]
\item
If $\varphi(a, a) = 1$
then $a \in f^{(s, s^\prime, t)}(T, \varphi)$ if and only if
$|{a^\prime \in T : \varphi(a^\prime, a) = 1}| \geq s$.
\item
If $\varphi(a, a) = \star$
then $a \in f^{(s, s^\prime, t)}(T, \varphi)$ if and only if
$|{a^\prime \in T : \varphi(a^\prime, a) = 1}| \geq s^\prime$.
\item
If $\varphi(a, a) = -1$
then $a \not\in f^{(s, s^\prime, t)}(T, \varphi)$ if and only if
$|{a^\prime \in T : \varphi(a^\prime, a) = -1}| \geq t$.
\end{itemize}
\setlist[itemize,1]{label=\textbullet}

In other words, if an individual qualifies or disqualifies themselves, then their social qualification is determined the same way as with the consent rule $f^{(s, t)}$. If an individual is indifferent about themselves, then they are socially qualified if and only if at least $s^\prime$ other individuals qualify them.

A special case of this rule is when $s^\prime = \lceil \frac{n + 1}{2} \rceil$ where $n = |N|$, i.e.\@ an individual who is indifferent about their own qualification is socially qualified if and only if a majority of the individuals qualify them. This special case is denoted as $f^{(s, \star, t)}$, and it is the basis of all the proofs in this section.

\begin{table}[!htb]
\begin{tabularx}{\textwidth}{
p{0.22\textwidth}
X
X
X
X
}
\hline & \multicolumn{4}{l}{$f^{(s, \star, t)}$ rules} \\
\cline { 2 - 5 } & $s=1$ & & $s \geq 2$ & \\
\cline { 2 - 5 } & $t=1$ & $t \geq 2$ & $t=1$ & $t \geq 2$ \\
\hline

% constructive

CGCAI &
NP-c &
NP-c &
NP-c &
NP-c \\

CGCDI &
NP-c &
NP-c &
NP-c &
NP-c \\

CGCPI &
? &
NP-c &
? &
NP-c \\

CGB / \$CGB &
NP-c &
NP-c &
NP-c &
NP-c \\

CGMB / \$CGMB &
P &
P &
P &
P \\

\hline
% destructive

DGCAI &
NP-c &
NP-c &
NP-c &
NP-c \\

DGCDI &
NP-c &
NP-c &
NP-c &
NP-c \\

DGCPI &
? &
? &
NP-c &
NP-c \\

DGB / \$DGB &
NP-c &
NP-c &
NP-c &
NP-c \\

DGMB / \$DGMB &
P &
P &
P &
P \\

\hline
% exact

EGCAI &
NP-c &
NP-c &
NP-c &
NP-c \\

EGCPI &
? &
? &
? &
? \\

EGB / \$EGB &
NP-c &
NP-c &
NP-c &
NP-c \\

EGMB / \$EGMB &
P &
P &
P &
P \\

\hline
% general

GCAI &
NP-c &
NP-c &
NP-c &
NP-c \\

GCDI &
NP-c &
NP-c &
NP-c &
NP-c \\

GCPI &
? &
NP-c &
NP-c &
NP-c \\

GB / \$GB &
NP-c &
NP-c &
NP-c &
NP-c \\

GMB / \$GMB &
P &
P &
P &
P \\

\hline
\end{tabularx}
\caption{
A summary of the complexity results for various group control and bribery problems when using the $f^{(s, \star, t)}$ rule with a ternary profile.
In the table, ``P'' stands for ``polynomial-time solvable'', ``NP-c'' stands for ``NP-complete'', and ``?'' means that the complexity of the problem is open.
}
\label{tab:ternary}
\end{table}

\tableref{tab:ternary} lists the known complexity results for various group control and bribery problems with the ternary profile extension and the $f^{(s, \star, t)}$ rule.

The results for CGCAI, CGCDI, CGCPI, CGB as well as DGCAI, DGCDI, DGCPI, DGB are all by \textcite[Section 6]{ERY20}. Their proofs are based on reductions from the \textsc{Restricted Exact Cover by 3-sets} problem.

Obviously, the NP-completeness results for CGB and DGB also extend to the priced versions \$CGB and \$DGB.

For the general and exact problems, we can obtain many NP-completeness results by simply extending the results from the constructive and destructive cases.

Both $f^{(s, \star, t)}$-EGCAI and -GCAI are NP-complete for all $s$ and $t$:

\begin{Corollary}
$f^{(s, \star, t)}$-\textsc{Exact Group Control by Adding Individuals} and \\ $f^{(s, \star, t)}$-\textsc{Group Control by Adding Individuals} are NP-complete for all values of $s$ and $t$.
\end{Corollary}

\begin{Proof}
This follows from the NP-completeness proof for $f^{(s, \star, t)}$-CGCAI by \textcite[Theorem 11]{ERY20}. In the proof, they have $A^+ = T \setminus \{d\}$ (i.e.\@ all individuals except the dummy must be socially qualified). We can easily turn this into an instance of EGCAI by setting $A^+ = T$ (i.e.\@ everyone including the dummy must be socially qualified) and letting everyone qualify the dummy.
\end{Proof}

Likewise, $f^{(s, \star, t)}$-GCDI is NP-complete for all $s$ and $t$:

\begin{Corollary}
$f^{(s, \star, t)}$-\textsc{Group Control by Deleting Individuals} is NP-complete for all values of $s$ and $t$.
\end{Corollary}

\begin{Proof}
This follows from the NP-completeness proof for $f^{(s, \star, t)}$-CGCDI by \textcite[Theorem 12]{ERY20}.
\end{Proof}

When $s \geq 2$ or $t \geq 2$, $f^{(s, \star, t)}$-GCPI is also NP-complete:

\begin{Corollary}
$f^{(s, \star, t)}$-\textsc{Group Control by Partitioning of Individuals} is NP-complete for all $s \geq 1$ and $t \geq 2$ as well as for all $s \geq 2$ and $t \geq 1$.
\end{Corollary}

\begin{Proof}
The NP-completeness for all $s \geq 1$ and $t \geq 2$ follows from the NP-completeness of $f^{(s, \star, t)}$-CGCPI \cite[23]{ERY20}. Similarly, the NP-completeness for all $s \geq 2$ and $t \geq 1$ follows from the NP-completeness of $f^{(s, \star, t)}$-DGCPI \cite[23]{ERY20}.
\end{Proof}

Both $f^{(s, \star, t)}$-EGB and -GB are NP-complete for all $s$ and $t$:

\begin{Corollary}
$f^{(s, \star, t)}$-\textsc{Exact Group Bribery} and $f^{(s, \star, t)}$-\textsc{Group Bribery} are NP-complete for all values of $s$ and $t$.
\end{Corollary}

\begin{Proof}
This follows from the NP-completeness proof for $f^{(s, \star, t)}$-CGB by \textcite[Theorem 15]{ERY20}. In the proof, they have $A^+ = N \setminus \{ a_H : H \in \mathcal{H} \}$. We can easily turn this into an instance of EGB by setting $A^+ = N$ (i.e.\@ everyone must be socially qualified) and letting everyone qualify the individuals in $\{ a_H : H \in \mathcal{H} \}$.
\end{Proof}

Finally, for the microbribery problems, it is easy to see that the polynomial-time algorithm by \textcite[Observation 2]{BBKL20} also works for ternary profiles. We simply determine for each individual in $A^+$ [resp.\@ $A^-$] what is the cheapest way to make them socially qualified [resp.\@ disqualified]. This computation can be done separately for each individual in polynomial time by looking only at their incoming qualifications. Hence, CGMB, \$CGMB, DGMB, \$DGMB, EGMB, \$EGMB, GMB, and \$GMB are all in P.

\clearpage

\section{Partial profiles (with missing information)}
\label{sec:partial}

In this section, we focus on group identification in the setting of partial profiles. The content of this section is based on the research done by \textcite{ERY17,R18}.

In some real-world applications, the full set of qualifications and disqualifications might not be known to us. For example, this could be the case when the number of individuals is extremely large, or when the individuals are not willing to reveal their full valuations. In scenarios like this, if we want to predict the result of the group identification process, we must do so based on only a part of the information.

Specifically, there are two main questions related to group identification in partial profiles: For which individuals is it even possible to be socially qualified if the missing information is filled, and which individuals are definitely socially qualified regardless of how the missing information is filled. In the following, we first provide formal definitions of partial profiles and the associated problems. We then give an overview of the known complexity results for each of these problems under different social rules.

A partial profile over $N$ is a function $\varphi : N \times N \rightarrow \{ -1, 1, \ast \}$. For $a,b \in N$, we write $\varphi(a,b) = \ast$ to denote that we do not know whether $a$ qualifies $b$ or not. This is not to be confused with a ternary profile where the individual themself is undecided whether they should qualify someone or not (see \secref{sec:ternary}). A partial profile $\varphi$ can be extended to a binary profile $\varphi^\prime$ by replacing each $\ast$-entry with either $1$ or $-1$. In that case, we call $\varphi^\prime$ an \textit{extension} of $\varphi$.

Recall that we use the term $r$-profile to refer to a binary profile where each individual qualifies exactly $r$ individuals (see \secref{sec:r_profiles}). Analogous to this definition, we now define a special type of partial profiles called $r$-partial profiles:

An $r$-partial profile over $N$ is a partial profile over $N$ where each individual $a \in N$ qualifies at most $r$ individuals. In this work, we only consider $r$-partial profiles that can be extended to $r$-profiles, i.e.\@ the $\ast$-entries can be replaced by $1$ or $-1$ in such a way that each individual qualifies exactly $r$ individuals. If some $r$-partial profile $\varphi$ can be extended to the $r$-profile $\varphi^\prime$, we call $\varphi^\prime$ an \textit{$r$-extension} of $\varphi$.

We now define the \textsc{Possibly Qualified Individuals} problem (PQI) and the \textsc{Necessarily Qualified Individuals} problem (NQI):

\bigskip

\renewcommand{\arraystretch}{1}

\begin{tabularx}{\textwidth}{lX}
\hline
\multicolumn{2}{l}{
$f$-\textsc{Possibly}/$f$-\textsc{Necessarily Qualified Individuals} ($f$-PQI/$f$-NQI)
} \\
\hline
\textbf{Given:} &
A 3-tuple $(N, \varphi, S)$ of a set $N$ of individuals, a partial profile $\varphi$ over $N$, and a nonempty subset $S \subseteq N$. \\
\textbf{$f$-PQI:} &
Is there an extension $\varphi^\prime$ of $\varphi$ such that $S \subseteq f(N, \varphi^\prime)$? \\
\textbf{$f$-NQI:} &
Does $S \subseteq f(N, \varphi^\prime)$ hold for every extension $\varphi^\prime$ of $\varphi$? \\
\hline
\end{tabularx}

\renewcommand{\arraystretch}{1.8}

Given a set $N$, a partial profile $\varphi$, and a subset $S \subseteq N$, the $f$-PQI problem asks whether it is possible to fill the $\ast$-entries of $\varphi$ in such a way that all individuals in $S$ are socially qualified with respect to the social rule $f$. The $f$-NQI problem asks whether the individuals in $S$ are always socially qualified with respect to $f$ regardless of how the $\ast$-entries of $\varphi$ are filled.

Both $f$-PQI and $f$-NQI can be restricted to $r$-partial profiles. In this case, we refer to them as $f$-$r$-PQI and $f$-$r$-NQI, respectively. In the restricted versions, we require $\varphi$ to be an $r$-partial profile and $\varphi^\prime$ to be an $r$-extension of $\varphi$. In other words, given a set $N$, an $r$-partial profile $\varphi$, and a subset $S \subseteq N$, the $f$-$r$-PQI problem asks whether it is possible to extend $\varphi$ to an $r$-profile $\varphi^\prime$ in such a way that all individuals in $S$ are socially qualified with respect to $f$ and $\varphi^\prime$. The $f$-$r$-NQI problem asks whether the individuals in $S$ are socially qualified with respect to $f$ for all $r$-extensions $\varphi^\prime$ of $\varphi$.

\tableref{tab:partial} lists the known complexity results for $r$-PQI, $r$-NQI, PQI, and NQI under the different social rules.

\renewcommand\cellalign{lc}

\begin{table}[!htb]
\begin{tabularx}{\textwidth}{
p{0.07\textwidth}
p{0.08\textwidth}
p{0.08\textwidth}
X
p{0.08\textwidth}
X
X
p{0.16\textwidth}
}
\hline & \multicolumn{6}{l}{Consent rules $f^{(s, t)}$} & \multicolumn{1}{c}{$f^{\text{CSR}} / f^{\text{LSR}}$} \\
\cline { 2 - 7 } & $s=1$ & & & $s \geq 2$ & & & \\
\cline { 2 - 7 } & $t=1$ & $t=2$ & $t \geq 3$ & $t=1$ & $t=2$ & $t \geq 3$ & \\
\hline

\multicolumn{1}{l|}{$r$-PQI} &
$\mathcal{O}(n^2)$ &
$\mathcal{O}(n^2)$ &
$\mathcal{O}(n^{r+t+2})$ &
$\mathcal{O}(n^3)$ &
$\mathcal{O}(n^{r+t+2})$ &
\multicolumn{1}{l|}{$\mathcal{O}(n^{r+t+2})$} &
\makecell{$r=1$: $\mathcal{O}(n^2)$ \\ $r \geq 2$: NP-c} \\

\cline { 2 - 8 }

\multicolumn{1}{l|}{$r$-NQI} &
\multicolumn{6}{c|}{$\mathcal{O}(n^2)$ ~} &
\makecell{$r=1$: $\mathcal{O}(n^2)$ \\ $r \geq 2$: $\mathcal{O}(n^4)$} \\

\cline { 2 - 8 }

\multicolumn{1}{l|}{PQI} &
\multicolumn{6}{c|}{$\mathcal{O}(n^2)$ ~} &
\multicolumn{1}{c}{$\mathcal{O}(n^2)$} \\

\cline { 2 - 8 }

\multicolumn{1}{l|}{NQI} &
\multicolumn{6}{c|}{$\mathcal{O}(n^2)$ ~} &
\multicolumn{1}{c}{$\mathcal{O}(n^2)$} \\

\hline
\end{tabularx}
\caption{
A summary of the complexity results for $r$-PQI, $r$-NQI, PQI and NQI. In the table, $n$ denotes the number of individuals and ``NP-c'' stands for ``NP-complete''.
}
\label{tab:partial}
\end{table}

\renewcommand\cellalign{lt}

For PQI and NQI, there exist simple $\mathcal{O}(n^2)$-time algorithms for all considered social rules \cite[Theorems 1 and 2]{ERY17}. The algorithms are based on the fact that a socially qualified individual is still socially qualified when someone who disqualifies them turns to qualify them. Hence, for the PQI problem, it suffices to check if the individuals in $S$ are socially qualified in the ``best case scenario'' (where many $\ast$-entries can be reset to $1$). Similarly, for the NQI problem, it suffices to check if the individuals in $S$ are socially qualified in the ``worst case scenario'' (where many $\ast$-entries can be reset to $-1$).

For all consent rules, $r$-NQI can be solved in time $\mathcal{O}(n^2)$ via a simple case analysis \cite[Theorem 3]{ERY17}. For the $f^{(1, 1)}$ and $f^{(1, 2)}$ rules, $r$-PQI can also be solved in time $\mathcal{O}(n^2)$ via a case analysis \cite[Theorem 4]{ERY17}. For the $f^{(2, 1)}$ rule, it can be solved in time $\mathcal{O}(n^3)$ by reducing it to \textsc{Maximum Flow} \cite[Theorem 4]{ERY17}. \textcite[Theorem 8.5]{R18} observes that the $\mathcal{O}(n^3)$ time algorithm can even be applied to all consent rules with $s \geq 2$ and $t=1$. In general, $f^{(s, t)}$-$r$-PQI can be solved in time $\mathcal{O}(n^{r+t+2})$ by guessing for each $a \in N$ with $\varphi(a, a) = \ast$ whether they qualify themselves or not and then applying the reduction to \textsc{Maximum Flow} \cite[Theorem 5]{ERY20}.

For the $f^{\text{CSR}}$ and the $f^{\text{LSR}}$ rule, $r$-PQI and $r$-NQI can be solved in time $\mathcal{O}(n^2)$ when $r=1$ via a simple case analysis \cite[Theorem 6]{ERY17}. For larger values of $r$, the $r$-PQI problem is NP-complete \cite[Theorems 8.10 and 8.11]{R18} while $r$-NQI can be solved in time $\mathcal{O}(n^4)$ \cite[Theorem 8.12]{R18}. The NP-completeness of $r$-PQI is shown by a reduction from the \textsc{Hamiltonian Path} problem, and the polynomial-time result for $r$-NQI is achieved via a more involved algorithm based on flow networks.

\clearpage

\section{Potential Parameters}
\label{sec:potential_params}

In this section, we briefly discuss potential parameters that could be used in algorithms for the problems presented in this work.

The most natural parameters for group control and bribery problems are the size of the target sets $|A^+|$ and $|A^-|$ as well as the budget $\ell$. For consent rules, there are also the quotas $s$ and $t$. Various combinations of these parameters have already been studied for different problems, as we outlined in \secref{sec:binary}. But there are still open questions, e.g.\@ whether $f^{\text{CSR}} / f^{\text{LSR}}$-GMB is XP (or even FPT) with respect to $|A^+| + |A^-|$. It is also not known whether the group control problems admit parameterized complexity results with respect to combinations of $\ell$, $s$ and $t$ in a similar way as it was shown for group bribery by \textcite{BBKL20}. Additionally, the question whether there is a parameter combination that can be used to obtain a polynomial-sized kernel is also open.

By interpreting problem instances as qualification graphs (see \secref{sec:basic_notation}), it is possible to derive additional parameters.
For example, this includes properties like the treewidth $\operatorname{tw}(G)$, the maximum degree $\Delta(G)$, and the minimum degree $\delta(G)$. It could be interesting to explore the similarities between qualification graphs of bounded degree and the restriction to $r$-profiles (see \secref{sec:r_profiles}). By using combinations of $\Delta$, $\ell$, $s$ and $t$ as parameters, \textcite{BBKL21} were able to obtain two FPT results for $f^{(s, t)}$-CGB. Furthermore, as pointed out by \textcite{BBKL20}, the (edge/vertex) distance of the qualification graph to a cluster graph, and the Kendall tau distance to a master (or central) profile could be useful parameters as well.

If we represent a given profile $\varphi$ as a binary matrix (see \exref{ex:one}), we can compute the hamming distances between the $i$-th matrix row and all other matrix rows (for some fixed $i$). When the hamming distances are small, we know that any individual in $N$ disagrees with the $i$-th individual only on a few positions. This fact could be used to solve certain group control and bribery problems faster.

In Sections \ref*{sec:constructive_param} and \ref*{sec:destructive_param}, we defined parameters $s^\ast$ and $t^\ast$ for instances of $f^{(s, t)}$-CGB/-\$CGB with $t=1$ and instances of $f^{(s, t)}$-DGB/-\$DGB with $s=1$, respectively. The parameters serve as a measure of how many qualifications (resp.\@ disqualifications) the individuals in the target set are missing, and how many choices we have to provide them. By a reduction from \textsc{Restricted Exact Cover by 3-sets}, we were able to show that the problems are NP-hard even for $s^\ast = 2$ and for $t^\ast = 2$, respectively. It would be interesting to investigate whether this still holds when the parameters are equal to $1$, or when they are used in combination with other parameters. Furthermore, for instances of $f^{(s, t)}$-CGCAI and -CGCDI as well as for their destructive counterparts, one could define similar parameters and check if these problems are also NP-hard when the parameters are small.

\clearpage

\section{Conclusion}
\label{sec:conclusion}

In this work, we have provided a comprehensive overview of manipulative attacks in group identification as well as group identification with partial profiles. We have listed all the currently known computational complexity results for the various problems and social rules.

For constructive and destructive group bribery with consent rules, we showed that the problems remain NP-hard even when each individual in the target set is already (dis)qualified by almost everyone and is only missing one further (dis)qualification. This implies that computing the solutions for these problem instances is generally difficult, and thus the corresponding social rules offer strong protection against bribery-based manipulative attacks (albeit not being fully immune to them).

We also extended the research on constructive and destructive group control problems in binary profiles to the general and exact cases. For this, we showed many immunity and NP-completeness results, but also a polynomial-time result for $f^{(2, 2)}$-GCDI, indicating that this rule should be avoided in scenarios where resilience against manipulative attacks is desired. Only a few cases in our analysis remain open. Furthermore, we generalized the FPT algorithms with respect to $|A^+|$ for $f^{(s, t)}$-CGCAI and -CGCDI to also work for $f^{(s, t)}$-GCAI and -GCDI with parameter $|A^+| + |A^-|$.

In the setting where the problems are restricted to $r$-profiles, we obtained many immunity, NP-completeness, and polynomial-time solvability results for CGCAI, building upon the existing research by \textcite{EY20} on the CGMB problem.

Regarding group control and bribery in ternary profiles, we extended the results by \textcite{ERY20} for the constructive and destructive problems to the the general and exact cases. We also studied the complexity of group microbribery in this setting, showing polynomial-time results for all corresponding problems. Although these polynomial-time results indicate weak protection against microbribery-based attacks, they could be useful for computing the margin of victory for certain individuals in a given instance.

For future research, the parameterized complexity of various problems could be studied, as we outlined in \secref{sec:potential_params}. Also, one could try to obtain complexity results for some of the open problems presented in this work. For many group control and bribery problems, the complexity when restricted to $r$-profiles could also be investigated, likely with similar approaches as shown in \secref{sec:r_profiles}.

One could also study group control and bribery problems in the context of partial profiles. In this setting, given a set $N$ and a partial profile $\varphi$, the attacker would try to manipulate the outcome of the group identification process in such a way that certain individuals are possibly (resp.\@ necessarily) socially qualified/disqualified.

Furthermore, it could be interesting to study problem variants of group control by deleting where we allow the attacker to delete individuals from $A^-$. This is motivated by the fact that, in many scenarios, the attacker would probably be satisfied if some or all of the individuals they want to turn socially disqualified simply got deleted instead.

Finally, one could consider problem instances of general and exact group control and bribery where $A^+$ and $A^-$ are both are nonempty, but for one of them the objective is already fulfilled, e.g.\@ all individuals in $A^+$ are already socially qualified initially. In particular, it would be interesting to see how this affects the immunity or hardness of these problems.

\clearpage
\nocite{*}
\printbibliography

\end{document}